\documentclass[printer]{aa}
\usepackage{graphicx}
\usepackage{textcomp}
\usepackage{txfonts}
\usepackage{xcolor}
\usepackage[utf8]{inputenc}
\usepackage{threeparttable}
\usepackage{multicol}

\usepackage{soul}

\title{The habitability of stagnant-lid Earths around dwarf stars}
  \author{Mareike Godolt
   \inst{1}
          \and
          Nicola Tosi \inst{1,2}
                    \and
          Barbara Stracke \inst{2}
          \and
          J.~Lee Grenfell \inst{2}
          \and Thomas Ruedas\inst{3,2}
          \and
          Tilman Spohn \inst{2}
          \and
          Heike Rauer \inst{1,2,4}
          }

   \institute{Zentrum f\"ur Astronomie und Astrophysik, Technische Universit\"at Berlin, Hardenbergstr. 36, 10623 Berlin, Germany\\
              \email{godolt@tu-berlin.de}
         \and  Institut f\"ur Planetenforschung, Deutsches Zentrum f\"ur Luft- und Raumfahrt, Rutherfordstraße 2, 12489 Berlin, Germany  
         \and  Leibniz-Institut f\"ur Evolutions- und Biodiversit\"atsforschung, Museum f\"ur Naturkunde,  Invalidenstr. 43,  10115 Berlin, Germany    
         \and  Institut f\"ur Geologische Wissenschaften, Freie Universit\"at Berlin, Malteserstr. 74--100, 12249 Berlin, Germany    }

\usepackage{natbib}
\usepackage{graphicx}

\abstract
    {The habitability of a planet depends on various factors, such as delivery of water during the formation, the co-evolution of the interior and the atmosphere, as well as the stellar irradiation which changes in time.
    }
    {Since an unknown number of rocky extrasolar planets may operate in a one-plate convective regime, i.e., without plate tectonics, we aim at understanding under which conditions planets in such a stagnant-lid regime may support habitable surface conditions. Understanding the interaction of the planetary interior and outgassing of volatiles in combination with the evolution of the host star is crucial to determine the potential habitability. M-dwarf stars in particular possess a high-luminosity pre-main sequence phase which endangers the habitability of planets around them via water loss. We therefore explore the potential of secondary outgassing from the planetary interior to rebuild a water reservoir allowing for habitability at a later stage.
    }
    {We compute the boundaries of the habitable zone around M, K, G, and F-dwarf stars using a 1D cloud-free radiative-convective climate model accounting for the outgassing history of CO$_2$ and H$_2$O from an interior evolution and outgassing model for different interior compositions and stellar luminosity evolutions.}
    {The outer edge of the habitable zone strongly depends on the amount of CO$_2$ outgassed from the interior, while the inner edge is mainly determined via the stellar irradiation, as soon as a 
    sufficiently large water reservoir has been outgassed. A build-up of a secondary surface and atmospheric water reservoir for planets around M-dwarf stars is possible even after severe water loss during the high luminosity pre-main sequence phase as long as some water has been retained within the mantle. For small mantle water reservoirs, between 62 and 125\,ppm, a time delay in outgassing from the interior permits such a secondary water reservoir build-up especially for early and mid M-dwarfs because their pre-main sequence lifetimes are shorter than the outgassing timescale. }
    {We show that Earth-like stagnant-lid planets allow for habitable surface conditions within a continuous habitable zone that is dependent on interior composition. Secondary outgassing from the interior may allow for habitability of planets around M-dwarf stars after severe water loss during the high-luminosity pre-main sequence phase by rebuilding a surface water reservoir.
    }
    
    \date{accepted 1st of March 2019}
    
\begin{document}
    
\maketitle

\section{Introduction}
\label{Introduction}

Nearly 4000 extrasolar planets have been found since their first detections in the 1990s. A major goal of exoplanetary research is to find and characterize potentially habitable planets. This requires a good understanding of the factors affecting their potential habitability. In an exoplanet context, habitability is usually defined by the presence of liquid water on the surface of the planet. This criterion puts constraints on the surface temperature and pressure and is used for the definition of the circumstellar habitable zone (HZ), which defines the range of orbital distances around a star where liquid water may be present on the surface, \citep[see e.g.][]{Hart1979,Kasting1993hz}. Whether or not water is really present critically depends on various factors, such as the atmospheric composition and mass, the water reservoir, and the evolution of these quantities.  

The most commonly used HZ boundaries computed by \citet{Kasting1993hz} and updated by \citet{Kopparapu2013} assume an Earth-like planet with an atmosphere composed of molecular nitrogen (N$_2$), carbon dioxide (CO$_2$) and water (H$_2$O) around M, K, G, and F-dwarf stars. The inner edge of the habitable zone is determined by evaluating the distance at which the entire water reservoir of 270~bar (one Earth ocean) would reside within the atmosphere (runaway greenhouse limit) or the distance at which the upper atmosphere becomes wet enough so that atmospheric escape would lead to the loss of one Earth ocean within the lifetime of the Earth, i.e., in 4.5\,Gyr. The outer edge is placed at the distance at which an Earth-like planet with an atmosphere providing a maximum greenhouse effect of CO$_2$ would have a surface temperature at the freezing point of water (273.15\,K). This maximum greenhouse effect occurs due to the competing processes of an increased absorption of thermal radiation and increased Rayleigh scattering of the stellar irradiation upon increasing the amount of CO$_2$ residing in the atmosphere. This approach is based on the assumption that a planet with an operating carbonate--silicate cycle, which is responsible for a stable climate on the Earth over long time spans \citep[see e.g.][]{Walker1981}, will regulate the amount of CO$_2$ in the atmosphere to obtain 
habitable conditions. However, the long-term carbonate--silicate cycle on the Earth involves plate tectonics, a tectonic regime which has been securely observed only for the Earth. Since the other terrestrial planets in the Solar System operate in a one-plate tectonic regime with a stagnant lid, one may assume that some of the rocky extrasolar planets may also possess a stagnant lid.

The habitability of planets without plate tectonics has been studied previously by e.g.~\citet{Noack2014}, \citet{Noack2017}, \citet{Tosi2017}, \citet{Dorn2018}, \cite{Valencia2018}, and \citet{Foley2018}. The latter two studies propose alternative mechanisms for carbonate cycling on stagnant-lid planets, whereas the others mainly focus on the outgassing of volatiles from the interior.
While most studies focus on the impact of outgassing (and cycling) of CO$_2$ from the planetary interior, in a previous paper \citep{Tosi2017} we investigated the habitability of stagnant-lid planets around the Sun accounting for the build-up of CO$_2$ and H$_2$O in the atmosphere via secondary outgassing from the interior. The study showed that stagnant-lid planets can in principle be habitable, though the habitable zone may be less extended due to limited outgassing for certain interior compositions. In addition, we found that an Earth-like ocean, i.e., with 270~bar, cannot result from secondary outgassing from the interior. 
In fact, because of the pressure-dependent solubility of water in basaltic melts, outgassing of H$_2$O strongly decreases as the pressure of the atmosphere increases over time. Nevertheless, depending on the assumed composition, a water reservoir of up to a few tens of bars can be outgassed from the interior \citep{Tosi2017}.

Building up a water reservoir via secondary outgassing may be especially important for the habitability of planets which experienced water loss during their early evolutionary stages, as proposed for planets around M dwarf stars. Stellar evolution models, e.g., by \citet{Baraffe2015}, clearly indicate that planets within the HZ around main-sequence M-dwarf stars have experienced a much higher stellar irradiation (up to 2--3 orders of magnitude) during the pre-main sequence phase of the star, which lasts longer for M-dwarf stars than for Sun-like stars. Consequences of this high luminosity pre-main sequence phase have been discussed e.g. by~\citet{Luger2015}, \citet{Ramirez2014}, \citet{Tian2015}, and \citet{Owen2016}.  This higher irradiation probably leads to much higher surface temperatures during the early evolution, leading to the evaporation of a potential surface water reservoir, photolysis of water molecules, and subsequent loss of hydrogen. Hence, rocky planets with a limited water reservoir may have lost their surface and atmospheric water during this early evolutionary stage. 

We investigate here whether secondary outgassing from the interior may rebuild a surface water reservoir after the high luminosity pre-main sequence phase. This will require that some water has been retained in the mantle despite an early phase of high stellar irradiation which probably led to severe atmospheric escape. Furthermore, higher irradiation during the early stages could lead to a long-term magma ocean as found for Earth-like planets around the Sun at distances smaller than 0.7\,au by \citet{Hamano2013}. Determining the duration of such a magma ocean phase and its consequences for the interior and atmospheric water reservoir would require a different modelling approach than the one used here and is beyond the scope of this paper. However, in order to account for different outcomes from such a magma ocean phase, we perform a parameter study using a variety of interior mantle water reservoirs. 
We base this study on the interior evolution and outgassing described in \citet{Tosi2017}. In the current paper we will study the evolution of the habitable zone boundaries of Earth-sized planets due to outgassing of CO$_2$ and H$_2$O and the luminosity evolution of their M, K, G and F-dwarf host stars. 

The paper is structured as follows: section \ref{comp_details} describes the interior evolution and outgassing, as well as the 1D radiative-convective climate model, which have been used to calculate the habitable zone boundaries, and motivates the modeled scenarios. Section \ref{results} shows and discusses the results. Here, we first show the interior and outgassing evolution in Sect.~\ref{Int_results} and then focus on the HZ evolution around the different (F, G, K, M) dwarf stars (Sect.~\ref{atmos_results}) for relatively large water reservoirs. In Sect.~\ref{Sec:M-Star_HZ} we then investigate the impact of the initial mantle water concentrations on the fate of the water reservoir of stagnant-lid planets around early to late M-dwarfs. After putting our results into a broader context in the discussions section (Sect. \ref{discussions}), we shortly summarize and conclude our study in Section \ref{summary}.

\section{Computational details}
\label{comp_details}

\subsection{Model description} \label{sec:model_description}

\begin{figure}[ht!]
    \centering
    \includegraphics[width=0.5\textwidth]{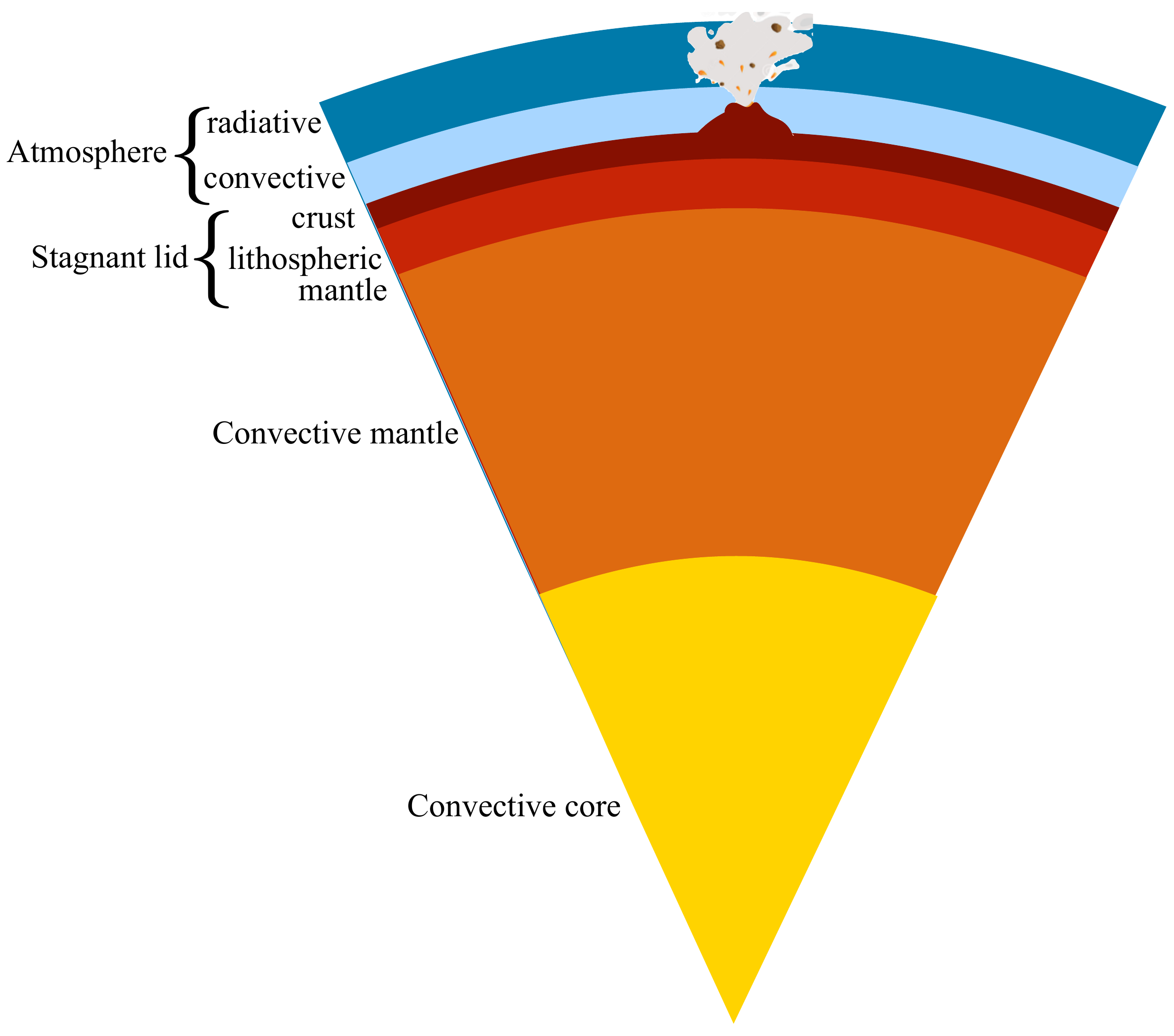}
    \caption{Sketch of the combined interior and atmosphere models.}
    \label{fig:model_sketch}
\end{figure}

Figure \ref{fig:model_sketch} shows a sketch of the modelling approach applied. For the interior we utilize a one-dimensional (1D) parameterized model of stagnant lid mantle convection. Partial melt generated in the convective mantle beneath the lithosphere buoyantly percolates upward building the crust. Carbon dioxide and water are outgassed from the melt that ultimately reaches the surface. The outgassed amount of these volatiles is then used in a 1D cloud-free radiative-convective climate model to calculate the boundaries of the habitable zone around different dwarf stars.\\

We follow the same approach as \citet{Tosi2017} to model the thermal evolution of the interior and the accompanying outgassing of a planet with the same mass, radius and core size as the Earth but operating in the stagnant lid mode of convection (i.e.~without plate tectonics) over 4.5 Gyr. Here we only present the main features of the model and refer the reader to the above study for a detailed description. 

We use a classic one-dimensional approach to parameterized mantle convection based on boundary layer theory \citep[e.g.][]{turcotte2002}. For an assumed initial temperature distribution, amount of long-lived radioactive elements, mantle water concentration and redox state, we solve the energy balance equations for the core, the convecting mantle, and the lithosphere. As shown in Fig. \ref{fig:model_sketch}, the latter comprises the uppermost part of the mantle (lithospheric mantle) and the compositionally distinct crust, which together constitute a mechanical unit that does not participate in the convection of the deeper interior due to its high viscosity but sits stably on top; because of this stability, which defines the key structural and dynamical difference from the lithosphere of our own Earth with its plate tectonics, it is also referred to as a ``stagnant lid''.   
Upon parameterizing the convective heat transfer, we employ a mantle viscosity based on diffusion creep of olivine dependent on temperature, pressure, and water concentration \citep{hirth2003}. In addition, the model accounts for partial melting, crust production, and partitioning of incompatible elements between mantle and crust (i.e., mantle depletion and crust enrichment). By comparing the mantle temperature with a peridotitic solidus dependent on water concentration \citep{katz2003}, we apply a model of accumulated fractional melting to extract radioactive elements and water from the mantle and partition them into the crust \citep{morschhauser2011}. In contrast, given the poor solubility of carbon in silicate minerals and its tendency to form separate phases, we apply a model of redox melting \citep{hirschmann2008,grott2011} under the assumption that the mantle is sufficiently reducing for carbon to be present in one of its elemental forms (i.e., graphite or diamond). 

The pressure at which partial melting occurs in our models is always lower than the limit of about 8 GPa above which basaltic melts are expected to be denser than the mantle residuum \citep[e.g.,][]{agee2008}. The melt produced is thus positively buoyant and tends to percolate upward from the source region through the lithosphere. Eventually, it forms new crust either by intrusion into already existing crust at depth or by extrusion at the surface, whereby the ratio of intrusive to extrusive volcanism is set to an intermediate value of 2.5 \citep[see e.g.][]{white2006}. The melt is enriched in water and CO$_2$. 
Whether or not these are ultimately outgassed into the atmosphere depends on whether the melt carrying them is erupted or intruded and on their solubility in erupted surface melts at the evolving pressure conditions of the atmosphere \citep{gaillard2014}. H$_2$O and CO$_2$ are therefore only released into the atmosphere if their concentration in surface melts is in excess of saturation according to solubility curves for basaltic melts \citep{newman2002}.

To calculate the boundaries of the habitable zone of a stagnant-lid Earth around M, K, G, and F-dwarfs, we applied a one-dimensional (1D), cloud-free, radiative-convective climate model, which has been described in detail by \citet{vonParis2010} and \citet{vonParis2015}, and is based on \citet{Kasting1984} and \citet{Segura2003}.
The radiative transfer is split into a stellar and a thermal wavelength regime. The short wavelength regime treats the absorption and scattering of stellar irradiation using a $\delta$-two-stream method including Rayleigh scattering coefficients following the approach of \citet{Allen1973} and four-term correlated-k exponential sums covering a wavelength regime from 273.5\,nm to 4.545$\mu$m. This wavelength coverage is optimized for solar irradiation. Especially for late M-dwarfs the cut-off at 4.545\,$\mu$m leads to non-negligible loss in incoming radiation of up to $\approx$ 5\%, see also \citet{Wunderlich2019}. Hence, HZ boundaries obtained with the models lie closer to the star than would be expected when accounting for this missing portion of irradiation. The long-wavelength regime treats the absorption by CO$_2$ and H$_2$O in the wavelength regime from 1 to 500\,$\mu$m via correlated-ks computed based on HITEMP 1995 \citep{Hitemp1995}. The ckd continuum \citep{ckd} as well as collision-induced absorption as described in \cite{Kasting1984} for CO$_2$ and N$_2$-N$_2$ as described in \citet{vonParis2013N2} are included.
Convection is treated by applying a convective adjustment when the adiabatic lapse exceeds the radiative lapse rate, including latent heat release from H$_2$O or CO$_2$ where applicable. The water mixing ratio profile ($C_{\mathrm{H}_2\mathrm{O}}$) is calculated from the temperature profile, the saturation vapor pressure ($p_\mathrm{sat}$), and by assuming a relative humidity ($RH$): $C_{\mathrm{H}_2\mathrm{O}}=RH\frac{p_\mathrm{sat}}{p}$, with $p$ the pressure of the atmosphere. By making use of our 1D climate model we estimate global, diurnal mean values, without accounting for effects such as slow planetary rotation or an interactive hydrological cycle. A discussion on the potential influence of 3D processes is given in Sect.~\ref{discussions}.

\subsection{Model setup}
We have run inverse climate calculations, similar to those carried out by \citet{Kasting1993hz} and \citet{Kopparapu2013}. Instead of specifying a stellar irradiation and computing the surface and atmospheric temperatures from this \citep[as done in][]{Tosi2017}, we have specified a temperature profile, for which we then determine the planetary albedo at the top of the atmosphere and the outgoing long-wave radiation. From this we estimate the incoming stellar irradiation ($S_{\mathrm{eff}}$) required to balance the outgoing radiation via $S_{\mathrm{eff}}=\tfrac{F_\mathrm{IR,up}}{F_\mathrm{S,net}}$, with $F_\mathrm{IR,up}$ the upwelling thermal (infrared) radiation and $F_\mathrm{S,net}$, the net stellar radiation.

\begin{figure}
    \centering
    \includegraphics[width=0.5\textwidth]{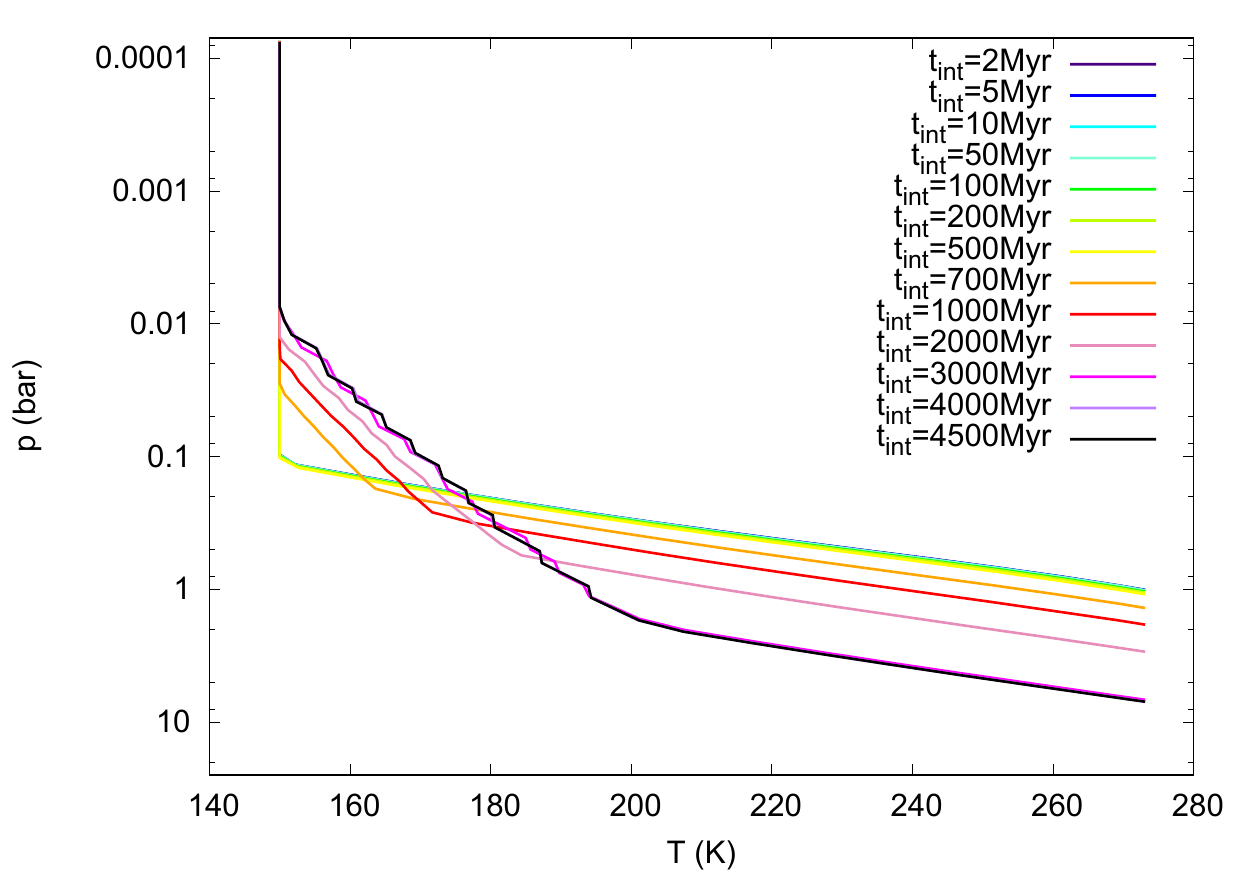}
    \includegraphics[width=0.5\textwidth]{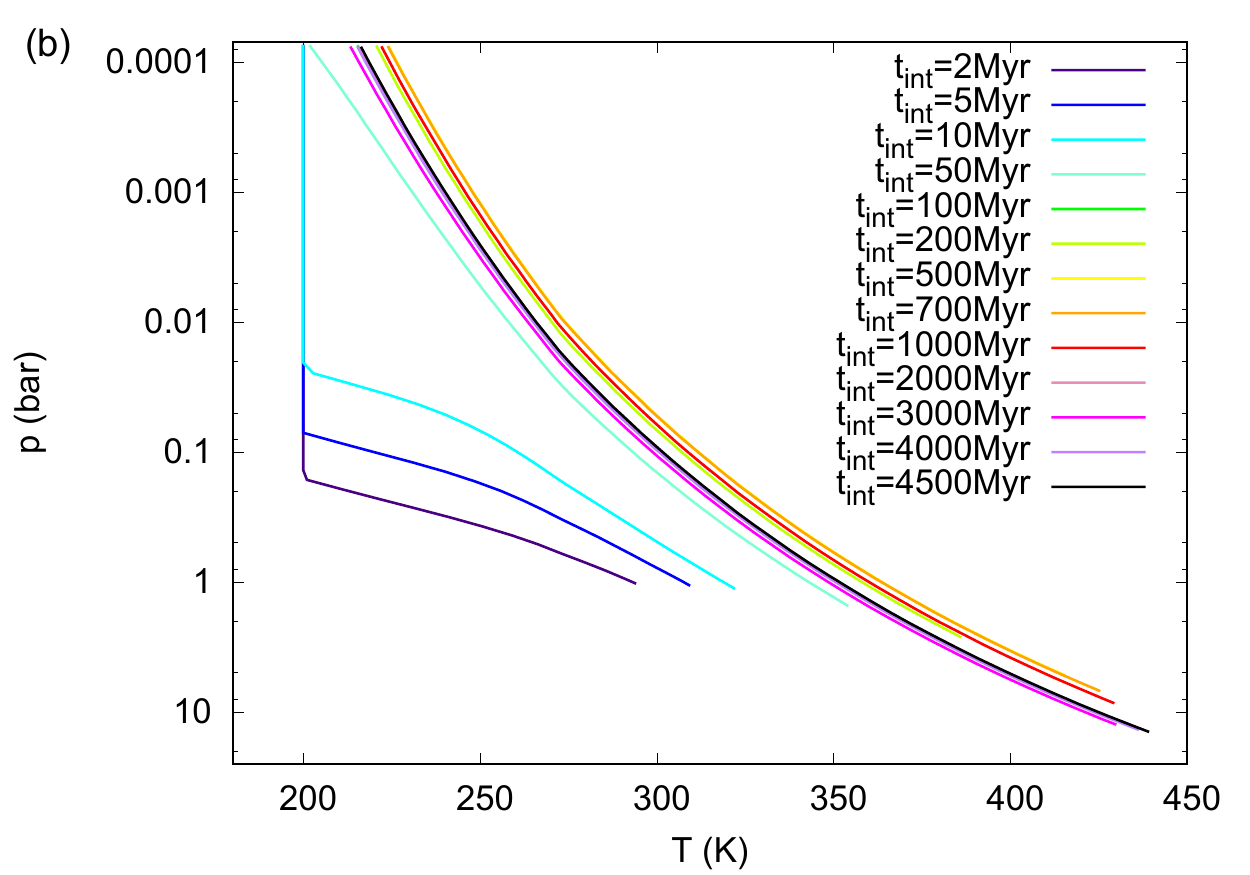}
    \caption{OHZ (a) and IHZ (b) temperature profile evolution for an initial water mantle concentration of 500\,ppm and at $\mathrm{IW}+0.5$}
    \label{fig:t_profiles}
\end{figure}

For the outer edge of the habitable zone we have set the surface temperatures to 273.15~K, the freezing point of water, and the stratospheric temperatures have been set to 150~K for all calculations.  Note that this is different from the approach of \citet{Kasting1993hz}, who used an albedo and irradiation dependent stratospheric temperature.
Using their expression would result in a range of stratospheric temperatures between $\approx$ 150 and 200\,K.  Varying the stratospheric temperature leads to a change in the upwelling thermal radiation, hence in $S_{eff}$. When using a stratospheric temperature independent of stellar irradiation, as we do here, the influence of the different stellar energy distributions only enters via the variable $F_\mathrm{S,net}$.
For the inner edge of the habitable zone we set the surface temperature to the value at which the entire outgassed water reservoir would be in the vapor phase for the assumption of phase equilibrium. This results in surface temperatures for the inner edge of the habitable zone that depend both on time and interior composition. We assume a stratospheric temperature of 200~K for these calculations, as in \cite{Kasting1993hz}. Under most conditions at the inner edge of the HZ, the upper atmospheric temperatures are mainly determined via convection (see Fig.~\ref{fig:t_profiles}), and much higher than 200\,K. For these cases the assumption on the stratospheric temperatures does not influence the results. We discuss the impact of the stratospheric temperature assumption on our results in Sect.~\ref{discussions}.   
With the above assumptions, habitable surface conditions can only be met if the outgassed water reservoir is larger than 6.11657\,mbar, i.e.~the saturation vapor pressure of water at 273.15~K. Figure \ref{fig:t_profiles} shows the temperature profiles at the OHZ and the IHZ boundaries along an evolution sequence for one of the scenarios described in the following section. The increase of the surface pressure with time caused by the outgassing of volatiles is clearly visible.

\subsection{Scenarios}\label{scenarios}

We study the impact of secondary outgassing associated with partial melting and volcanism following accretion and a possible magma ocean phase. \cite{Raymond2007} showed that the accretion of planets around M-dwarf stars is faster than around more massive stars, which has also been calculated by \cite{Lissauer2007} and \cite{Ida2005}. We therefore assume that the accretion of a planet  around an M-dwarf is very fast (1\,Myr), while for the other stellar cases we assume an accretion time of 20\,Myr.
This corresponds to the lower limit in accretion times of \citet{Raymond2007} for planets around stars with masses greater than $0.8 M_\mathrm{Sun}$.
After accretion, the planet may reside in a magma ocean stage, which we assume to last 1\,Myr, following the results by e.g.~\cite{Spohn1991} and \citet{Lebrun2013}.
For high stellar irradiation, the planet may be trapped in a long-term magma ocean stage, which may last from several millions of years up to billions of years \citep[see e.g.,][]{Hamano2013, Nikolaou2019}. For planets in a long-term magma ocean stage, \citet{Hamano2013} suggested very low water concentrations in the mantle at the end of the magma ocean stage because the planet can only exit this stage via severe atmospheric water loss because thick, water-rich atmospheres show a thermal blanketing effect suppressing the cooling of the planet via the Planck feedback. 
Their estimates of interior water reservoirs for such cases correspond to our lower limit of the initial mantle water concentration of 34\,ppm.  
We nevertheless assume in our calculations that the magma ocean has a short duration. This allows the high luminosity pre-main sequence phase of the M-dwarfs to have a larger impact on the HZ evolution. Any CO$_2$ and H$_2$O outgassed during the magma ocean phase is neglected in our calculations, i.e., we start our outgassing calculations with a 1\,bar N$_2$ atmosphere. Neglecting any atmospheric H$_2$O and CO$_2$ from the magma ocean phase results in a very narrow HZ at the beginning of our calculations, which is then only extended by secondary outgassing. This facilitates the evaluation of the potential of secondary outgassing from a stagnant-lid planet to form a HZ. 

To compute the HZ boundaries depending on the thermal and outgassing evolution of our Earth-sized stagnant-lid planets, we assume different interior compositions. For the outgassing of CO$_2$, the mantle oxygen fugacity ($f_{\mathrm{O_2}}$) is important \citep{Tosi2017}; it can be thought of as an effective partial pressure of oxygen in the mantle and is a measure of how oxidizing or reducing the mantle is. A mantle with a higher oxygen fugacity will allow a larger fraction of the carbon partitioned into the melt to be in oxidized form and can therefore also release more CO$_2$ into the atmosphere. We vary $f_{\mathrm{O_2}}$ between one log$_{10}$-unit below the iron--w\"ustite (IW) buffer and one log$_{10}$-unit above it, i.e., from $\mathrm{IW}-1$ to $\mathrm{IW}+1$. The IW buffer essentially defines the oxygen fugacity at which iron (Fe) and w\"ustite (FeO) are in thermodynamic equilibrium. The chosen values are two or three orders of magnitude lower than average values for common terrestrial mantle rock and hence correspond to the absence of plate tectonics and an Earth-like water cycle.

For the initial mantle water concentration, we consider values of 34, 62, 125, 250, and 500\,ppm for planets around M-dwarf stars, while for F, G, and K-dwarf stars, we only discuss the impact of an initial mantle water concentration of 500\,ppm \citep[the interior modelling results for 250\,ppm and 500\,ppm have already been discussed in detail by][] {Tosi2017}. We consider a range of initial mantle water concentrations especially for the planets around the M-dwarfs, since the long pre-main sequence phase may cause a long-term magma ocean phase which can lead to substantial water loss and a drier planetary interior (see also Sec.~\ref{Introduction}). At the lower end, we use 34\,ppm since \cite{Hamano2013} find an interior mantle water reservoir of 0.1 Earth oceans ($\approx$34\,ppm) after a planet has exited the magma ocean phase via water loss to space. Mechanisms that may lead to a larger fraction of volatiles trapped in the interior upon magma ocean crystallization have however also been discussed, e.g., by \citet{Hier-Majumder2017}. 

Furthermore, for the interior calculations we assume an initial mantle temperature of 1700\,K \citep[as in the reference model of][]{Tosi2017} and a surface temperature of 294 K. As discussed in \citet{Tosi2017}, differences in the initial mantle temperature tend to be rapidly erased because of the strong temperature dependence of the viscosity. Also, considering a constant surface temperature has a minimal impact on the melt production and outgassing for the range of mantle water concentrations and oxidation states considered in this work.

For the mantle radiogenic heating, we considered bulk silicate Earth concentrations of Uranium (U), Thorium (Th) and Potassium (K) according to \citet{mcdonough1995} (i.e.~20 ppb U, 80 ppb Th, and 240 ppm K). Variations of such concentrations, which we do not consider in this study, would affect the amount and timing of melt production and, in turn, of outgassing: a higher internal heating would facilitate earlier mantle melting with subsequently earlier and more extensive outgassing. For a given oxygen fugacity and initial water concentration in the mantle, a larger amount of CO$_2$ would be thus outgassed. However, because of the high solubility of water in surface melts (see Sec. \ref{sec:model_description}), a rapid increase of the CO$_2$ partial pressure would be accompanied by an early reduction of water outgassing. As a consequence, the final partial pressure of water in the atmosphere would be buffered to a nearly constant value, weakly dependent on the internal heat production.

We also assume that no primary crust generated by magma ocean solidification \citep[as, for example, in the case of the Moon, see e.g.][]{warren1985} is present at the beginning of the evolution, and set the initial stagnant lid thickness to 50 km. This choice only affects the earliest stages of the interior evolution. The lid thickness rapidly converges to a physically consistent value dictated by the internal energy budget. Simulations carried out using initial lid thicknesses between 25 and 150 km lead to differences in the main diagnostic quantities (see Sec.~\ref{Int_results}) of only up  to  few percent over the evolution with respect to those presented here.
In all cases, we compute the interior and outgassing evolution over a time span of 4.5 Gyr. 

We model planetary atmospheres that are composed of CO$_2$ and H$_2$O evolving with outgassing from the interior, and 1 bar N$_2$, which acts as a background gas. We furthermore assume a surface albedo of 0.22, and a relative humidity of 1. For the irradiation by the different host stars we account for the impact of different stellar spectral energy distributions as well as the luminosity evolution. We consider spectral energy distributions from the F-type dwarf star $\sigma$ Bootis, the Sun as an example for a G-type star, $\epsilon$ Eridani for a sample K-dwarf, and AD Leo as M-dwarf star. These spectral energy distributions are representative of the current evolutionary stage of these stars, with the Sun being the oldest while the other have ages between 25--300\,Myr to 1.7\,Gyr \citep[see e.g.][]{Shkolnik2009,Mamajek2008,Decin2003}. 
Details about the spectra can be found e.g.~in \cite{Kitzmann2010}. For the calculations of the M-star HZ boundaries in Sect.~\ref{Sec:M-Star_HZ} we have employed Planck curves for the stellar irradiation of M4 and M6-stars as in \cite{Rauer2011}, with effective temperatures of 3100\,K (M4), and 2600\,K (M6), being similar to the effective temperatures of Proxima Centauri and TRAPPIST-1.

\begin{figure}[h!]
\centering
\includegraphics[width=0.45\textwidth]{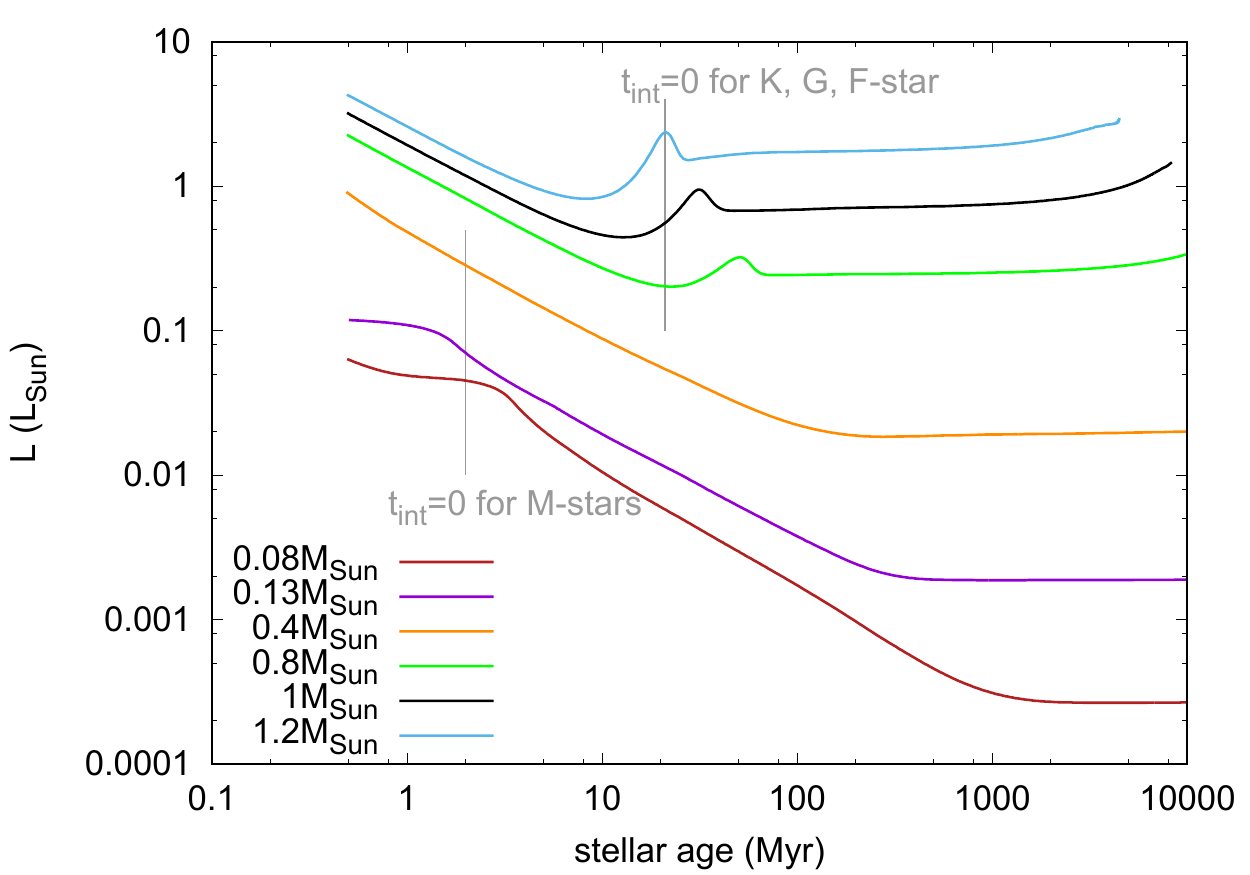}
\caption{Stellar luminosity evolutions used, following \citet{Baraffe2015}.}
\label{fig:stellar_luminosity_evolution}
\end{figure}

In addition to the spectral energy distribution, we need to assume a stellar evolution to calculate the evolution of the habitable zone boundaries over time. We use the stellar evolution models by \cite{Baraffe2015} to account for the change in stellar luminosity in time (see Fig.~\ref{fig:stellar_luminosity_evolution}). The stellar parameters used are summarized in Tab.~\ref{tab:stars}.

\begin{table}
 \caption{Stellar parameters}
     \label{tab:stars}
    \centering
    \begin{threeparttable}
    \begin{tabular}{c|c|c|c}
         Star& Stellar type & $T_\mathrm{eff}$ (K) & $M_\mathrm{Star}$ ($M_\mathrm{Sun}$) \\ \hline
         M6 & M6V & 2600 & 0.08\\ 
         M4 & M4V & 3100 & 0.13\\
         AD Leo & M3.5V\footnotemark[1] & 3390\footnotemark[0] & 0.42\footnotemark[1]\\
         $\epsilon$ Eridani & K2V\footnotemark[2]  &5039\footnotemark[2] & 0.82\footnotemark[2]\\
         Sun& G2V& 5777 & 1\\
         $\sigma$ Bootis & F2V\footnotemark[3]& 6594\footnotemark[3] & 1.194\footnotemark[3]\\
          \end{tabular}
    \begin{tablenotes}
    \begin{multicols}{2}
    \begin{small}
     \item[0]{\citet{Rojas-Ayala2012}}
\item[1]{\citet{Reiners2009}}
    \item[2]{\citet{Baines2012}}
    \item[3]{\citet{Boyajian2012}} 
    \end{small}
    \end{multicols}
    \end{tablenotes}
\end{threeparttable}
\end{table}
Figure \ref{fig:stellar_luminosity_evolution} clearly shows the long pre-main sequence phase of the M-Stars. For the $0.4 M_\mathrm{Sun}$ star this phase lasts about 100--200\,Myr, for the $0.13 M_\mathrm{Sun}$ star it lasts up to about 200--300\,Myr, while for the very low mass M-dwarf it lasts for more than 1000\,Myr. For the other dwarf stars with higher masses the pre-main sequence phase is much shorter. For them, however, a luminosity evolution during the main sequence phase exists. This luminosity increase with time has led to the famous faint young sun paradox for the Earth and Mars, which has been investigated by several groups \citep[see e.g.][]{Feulner2012,vonParis2013N2,Wolf2013, Kunze2014,Ramirez2014_Mars,Wordsworth2017}.

While the interior evolution and outgassing is obtained from time-dependent modelling, we perform time slice calculations for the atmospheres at $t_{int}$= 2, 5, 10, 50, 100, 200, 500, 700, 1000, 2000, 3000, 4000, and 4500\,Myr, hence at different timesteps of the interior evolution. We present the results for the interior evolution time ($t_{int}$) modeled, which starts at the end of the magma ocean phase, see Fig.~\ref{fig:stellar_luminosity_evolution}.

\section{Results}    
\label{results}

This section discusses the results of our interior evolution modelling (Sec. 3.1), and the habitable zone boundaries as found by the atmosphere modelling using the outgassing from the interior and the stellar evolution of the host stars in Sec.~\ref{atmos_results} for relatively high initial water concentrations in the mantle. In Section \ref{Sec:M-Star_HZ} we then explore the influence of the long pre-main sequence phase of M-dwarfs upon the habitability of stagnant-lid planets assuming different interior water reservoirs.

\subsection{Interior modelling results}
\label{Int_results}

\begin{figure}[!ht]
\begin{center}
    \includegraphics[width=0.45\textwidth]{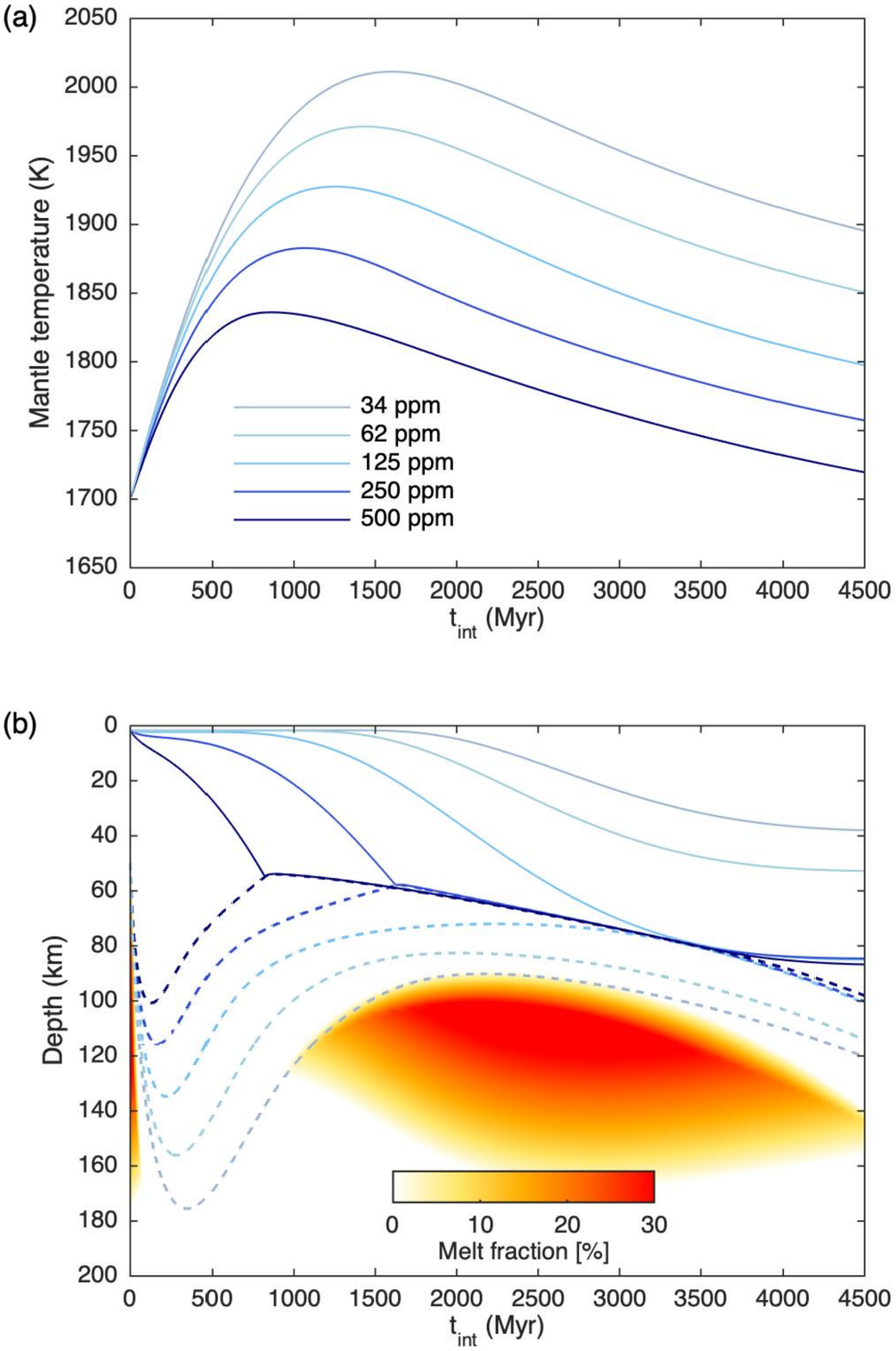} 
       \caption{a: Evolution of the mantle temperature for different water concentrations and an oxygen fugacity corresponding to the IW buffer. b: Corresponding evolution of the crustal thickness (solid lines) and  stagnant lid thickness (dashed lines). The coloured area denotes the distribution of the melt zone and melt fraction for the case with an initial water concentration of 62\,ppm.}
    \label{fig:interior_evolution}
        \end{center}
     \end{figure}

We ran a series of interior evolution models varying the initial water concentration of the mantle and the mantle oxidation state for a range of fixed (and not evolving) oxygen fugacities. Fig.~\ref{fig:interior_evolution} shows the evolution of the mantle temperature (a) and of the crustal and stagnant lid thickness (respectively solid and dashed lines in panel b) for different initial water concentrations in the mantle and an oxygen fugacity corresponding to the IW buffer. The time axis corresponds to the interior evolution time (t$_{int}$) which starts at the end of the magma ocean phase. Additionally, the evolution of the distribution of the melt fraction for an initial water concentration of 62\,ppm is shown in panel b. We show the results for 62\,ppm initial mantle water concentration as an example for an evolution of a relatively dry interior. The evolution of a wetter interior with 500\,ppm has been shown and discussed in \citet{Tosi2017}. During the early evolution, the mantle temperature increases largely due to radiogenic heating coupled with inefficient heat loss through the stagnant lid. The maximum temperature reached as well as the time span over which the mantle heats up increase with decreasing initial mantle water concentration.
This is due to the increase in the mantle viscosity upon lowering the water concentration, which in turn slows down convection and hence mantle cooling.  

Since the mantle solidus temperature increases with decreasing mantle water concentration, a comparatively hot mantle is needed for partial melting, crust formation, and outgassing if the initial mantle water concentration is low (34 to 62\,ppm).
For low initial water concentrations (e.g.~34 and 62\,ppm), after a short adjustment phase during which 1--2\,km crust are built, the bulk of the crust needs as long as $\sim$1000\,Myr to form (see partial melt zone in Fig.~\ref{fig:interior_evolution}b). Consequently, for these drier cases, the overall amount of crust produced tends to be relatively small compared to cases with larger initial water concentrations. This lower crust growth reduces both the reservoir from which the outgassing of volatiles can take place and the time span over which this can occur.
The cases with high water concentration (250 ppm and 500 ppm) are characterized by the rapid production of a large amount of partial melt, which causes the crust to quickly grow as thick as the stagnant lid. As soon as this condition is met (see e.g. $t_{int}\approx 750$ Myr for the case with initial water concentration of 500 ppm), crustal erosion begins and continues nearly until the end of the evolution. While in principle the crust could grow thicker according to the amount of melt that is continuously produced, it does not do this in our calculations because its base reaches the sub-lithospheric region where active convection recycles it back into the mantle. Therefore the growth of the crust is limited by the evolution of the stagnant-lid whose thickness increases upon mantle cooling (see also \citet{Tosi2017} for a detailed description of these cases).

\begin{figure}[ht!]
\begin{center}
    \includegraphics[width=0.45\textwidth]{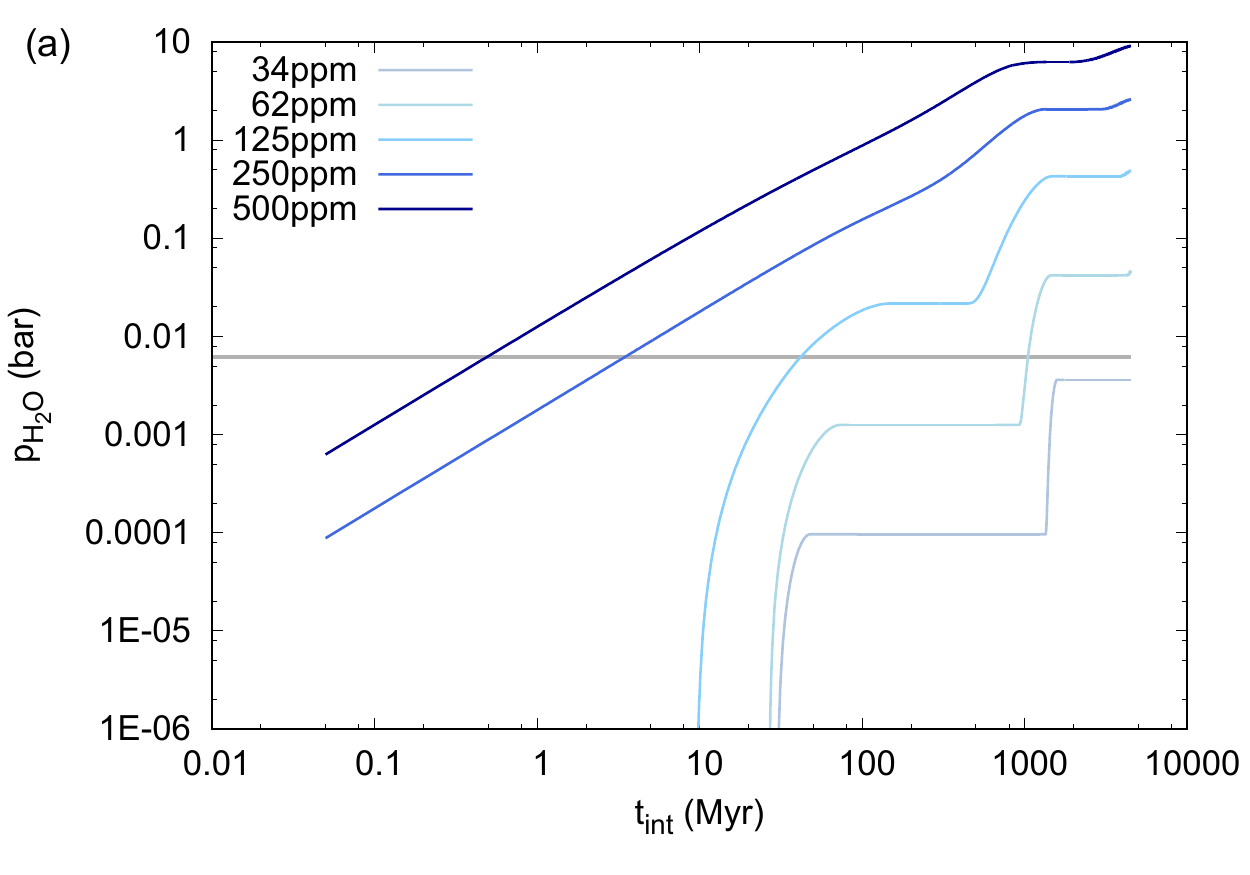}
    \includegraphics[width=0.45\textwidth]{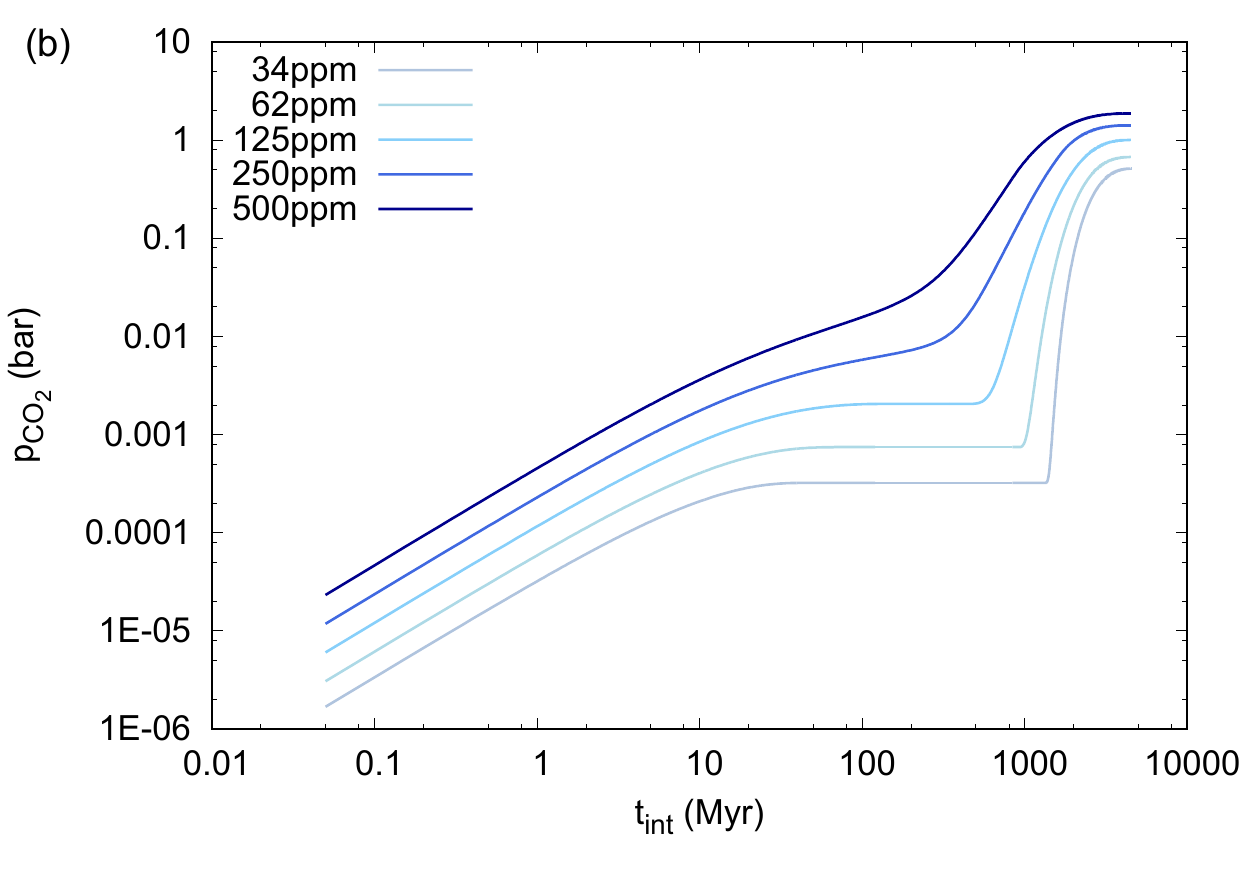}
        \caption{Outgassing of H$_2$O (a) and CO$_2$ (b) for the same cases shown in Fig.~\ref{fig:interior_evolution}. The horizontal grey line in the top panel corresponds to 6.12\,mbar, i.e. the saturation vapor pressure of water at 273.15\,K.}
          \label{fig_h2o_CO2_out_IW0}
        \end{center}
      \end{figure}

Figure \ref{fig_h2o_CO2_out_IW0} shows the outgassing of H$_2$O and CO$_2$ for the same cases shown in Fig. \ref{fig:interior_evolution}. As expected, the amount of water outgassed from the interior increases for increasing initial water concentrations. For initial water concentrations of 250 and 500\,ppm,  the outgassing is rather continuous until it vanishes between 500 and 1000\,Myr when surface melts start to be undersaturated in water \citep[see also][]{Tosi2017}. A small amount of water can be outgassed again after 4000\,Myr due to the fact that for the low melt fractions occurring near the end of the evolution, surface melts, albeit small in volume, tend to be highly enriched in water and hence supersaturated. For initial water concentrations below 250\,ppm, the outgassing of water follows multiple steps. Some outgassing occurs during the first 10 to 100\,Myr. After a quiet period during which no partial melting (see Fig.~\ref{fig:interior_evolution}) and thus no outgassing takes place (see Fig. \ref{fig_h2o_CO2_out_IW0}), a second period of outgassing is observed. This second period occurs later and is shorter for lower initial water mantle concentration. For initial mantle water concentrations of 62\,ppm and higher, outgassing of H$_2$O re-occurs during later stages.

A similar step-wise behaviour for the outgassing is also obtained for CO$_2$. For high initial mantle concentrations of water, more CO$_2$ is outgassed from the interior at a given oxygen fugacity due to the decrease in the solidus temperature for increasing water concentration, which allows for the production of more partial melt. 

From the outgassing evolution of water in Fig.~\ref{fig_h2o_CO2_out_IW0}a it is apparent that an Earth-like stagnant-lid planet with only 34~ppm water initially cannot produce habitable surface conditions via outgassing, since the resulting water reservoir is too small, i.e.~ below 6.12\,mbar, as indicated by the grey horizontal line. The entire outgassed water reservoir would reside within the atmosphere already at 273.15\,K, assuming phase equilibrium. 
Additional model calculations have shown that for an oxygen fugacity at the iron--w\"ustite buffer (IW) and an initial mantle water concentration of 38.6\,ppm, 6.12\,mbar of H$_2$O would be outgassed. The required initial mantle water concentration to reach 6.12\,mbar depends on the oxygen fugacity, which can be better understood by evaluating the outgassing evolution for different values of $f_{O_2}$.

\begin{figure}[ht!]

\begin{center}
\includegraphics[width=0.45\textwidth]{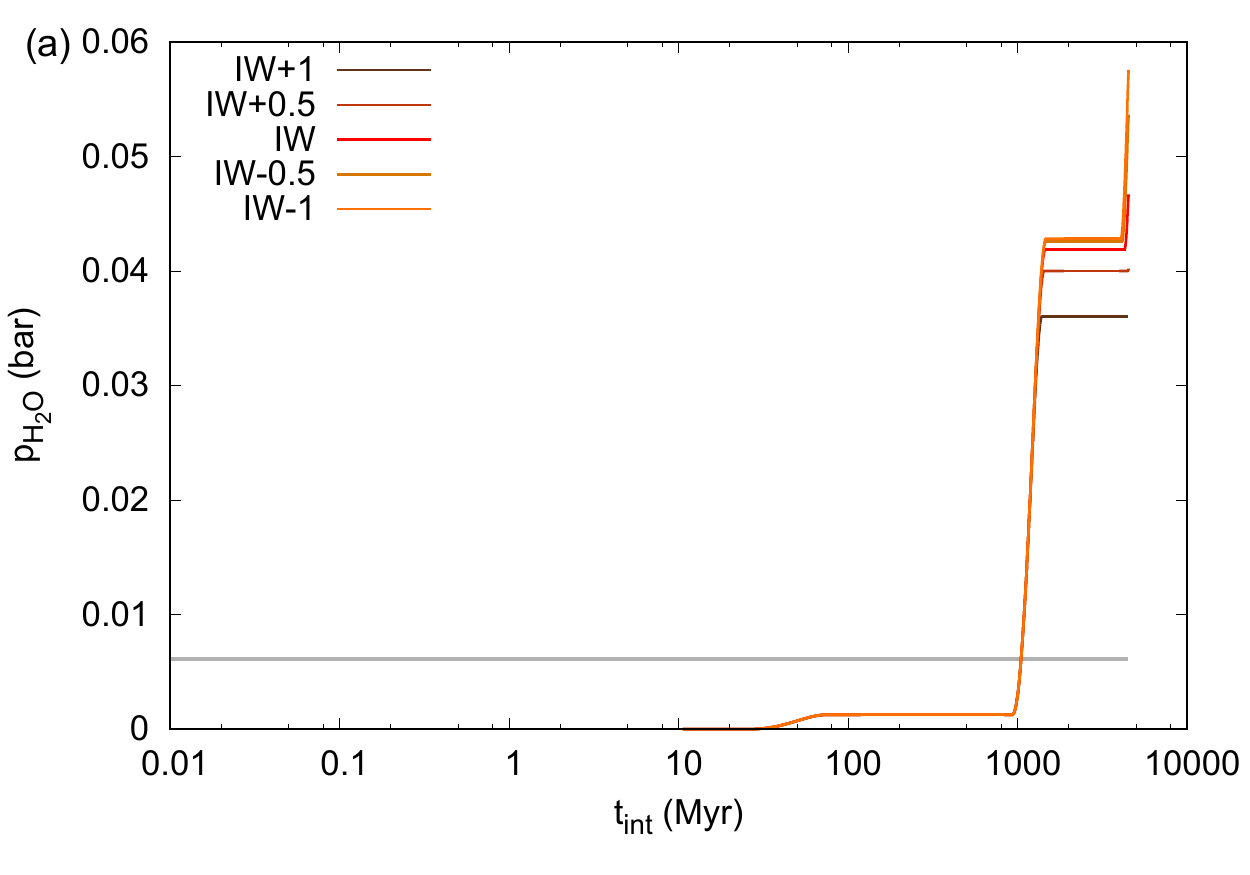}
\includegraphics[width=0.45\textwidth]{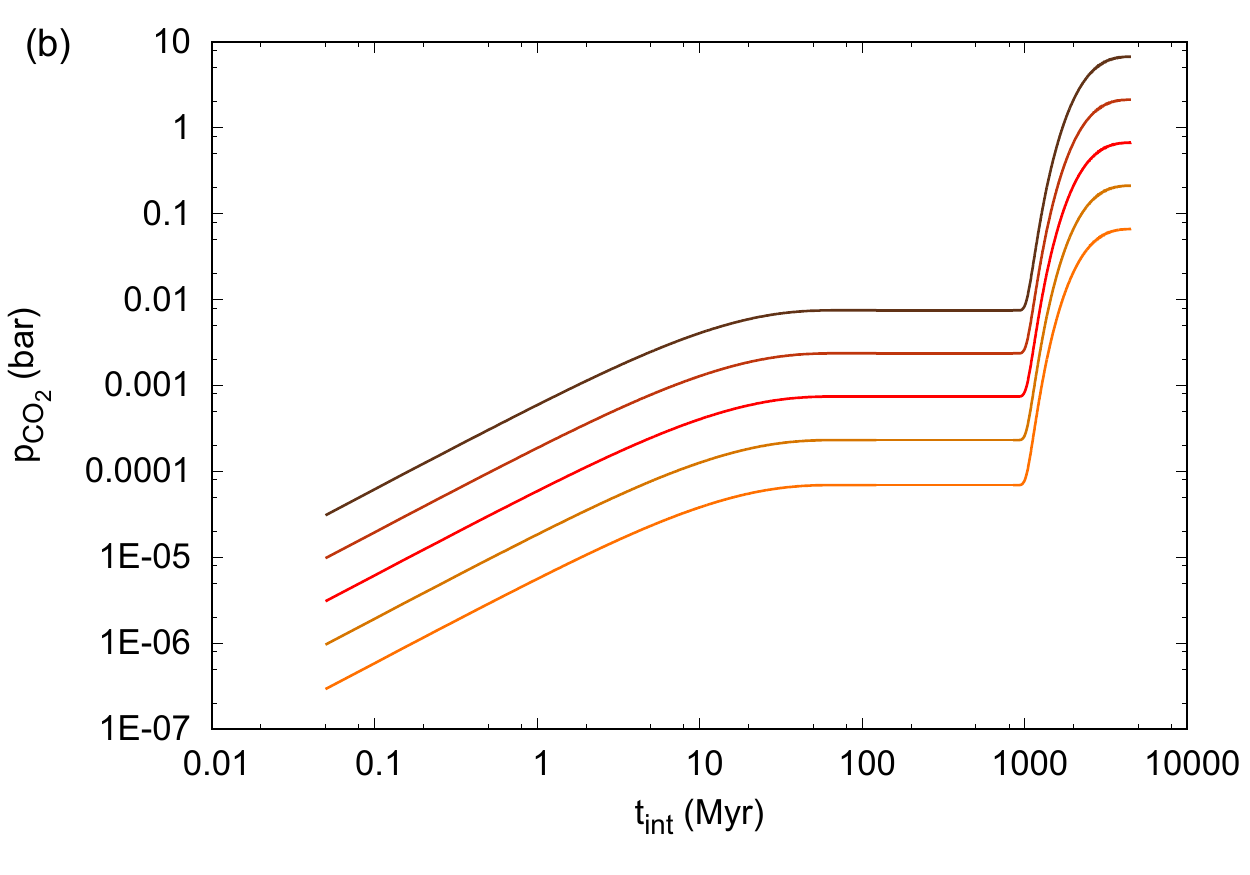}
 \end{center}
\caption{Outgassing evolution of H$_2$O (a) and CO$_2$ (b) for an initial mantle water concentration of 62\,ppm water and different oxygen fugacities. The grey horizontal line indicates the saturation vapor pressure of water at 273.15\,K. Note that the vertical axis of panel a is linear while it is logarithmic in panel b and in Fig.~\ref{fig_h2o_CO2_out_IW0}.}
\label{fig:out_H62}
\end{figure}

Figure \ref{fig:out_H62} shows the outgassing evolution of H$_2$O and CO$_2$ for an initial mantle water concentration of 62\,ppm for different oxygen fugacities.  For the outgassing of water, the impact of the outgassed CO$_2$ becomes visible from about 1300\,Myr, after the second outgassing phase. The increasing amount of CO$_2$ in the atmosphere increases atmospheric pressure and with this the solubility of H$_2$O in the melt, which leads to less outgassing, as discussed in \citet{Tosi2017}. Hence, the scenarios with the lowest CO$_2$ outgassing ($\mathrm{IW}-1$) have the highest amount of water outgassed from the interior. Overall the outgassing of water for this initial water concentration is very low. With a potential atmospheric and oceanic water reservoir of only about 0.05\,bar, this is more than 2000 times smaller than Earth's ocean reservoir of 270\,bar, and only about five times the amount of water within the Earth's atmosphere, which has a volume mixing ratio of about 1$\%$ on average. 

The amount of CO$_2$ increases with oxygen fugacity. Its stepwise increase with time is due to the stepwise melt production, as explained above. For this low initial water concentration, the amount of CO$_2$ outgassed from the interior can be as large as 6.7\,bar for the highest oxygen fugacity assumed here ($\mathrm{IW}+1$). For an initial water concentration of 500\,ppm, the final amount of CO$_2$ outgassed from the interior for $\mathrm{IW}+1$ is 19.1\,bar \citep[see][]{Tosi2017}.\\

\subsection{Evolution of the habitable zone boundaries of stagnant-lid planets around different dwarf stars }\label{atmos_results}

    \begin{figure*}[ht!]\begin{center}
    \includegraphics[width=0.4\textwidth]{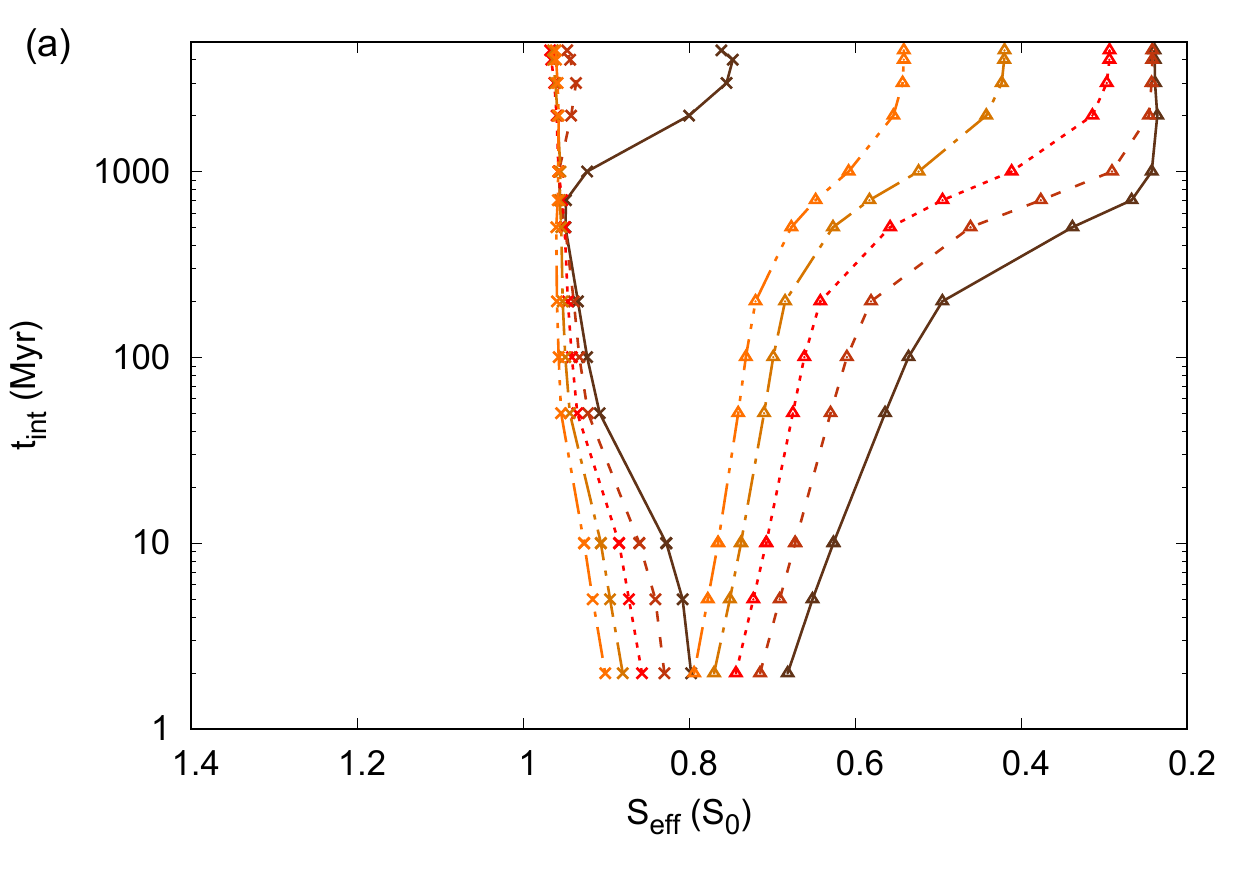}
    \includegraphics[width=0.4\textwidth]{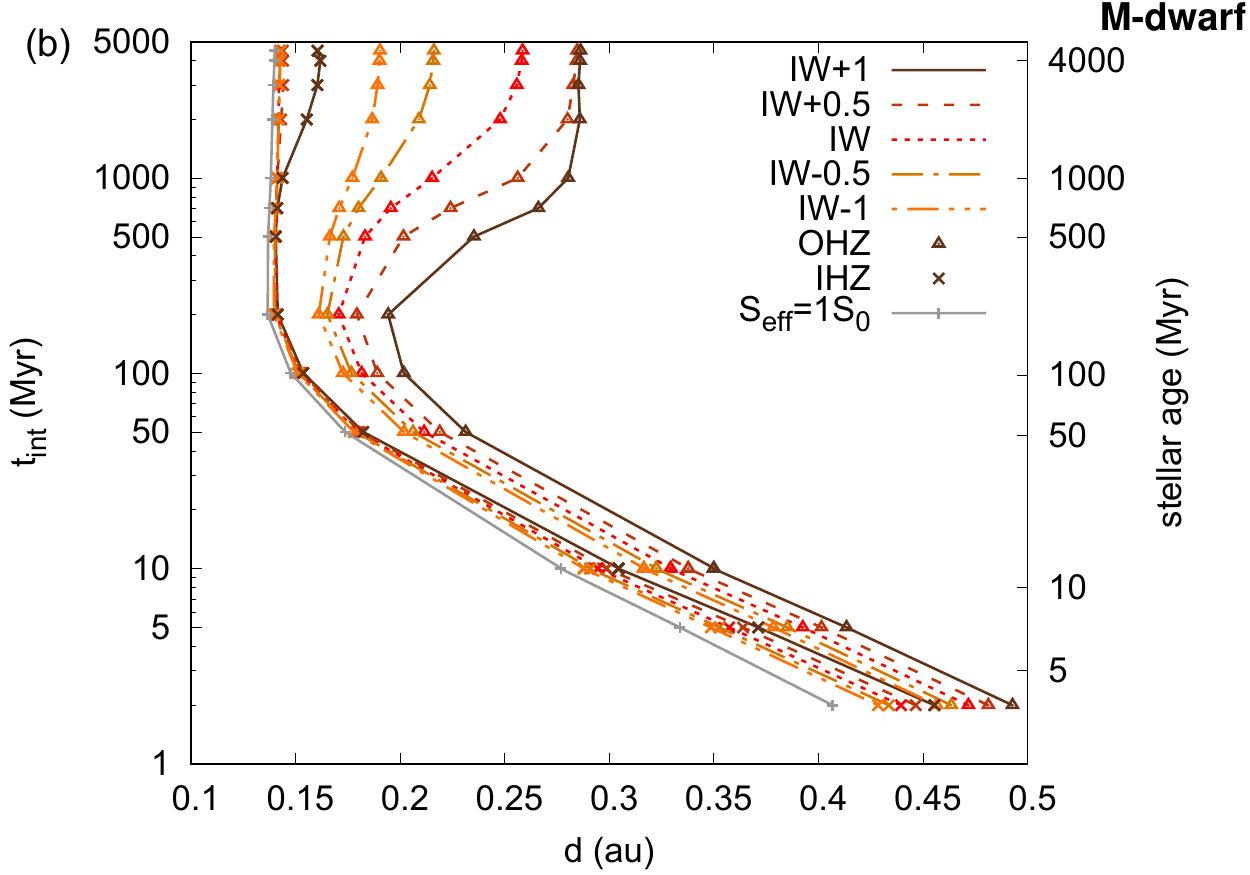}
    \includegraphics[width=0.4\textwidth]{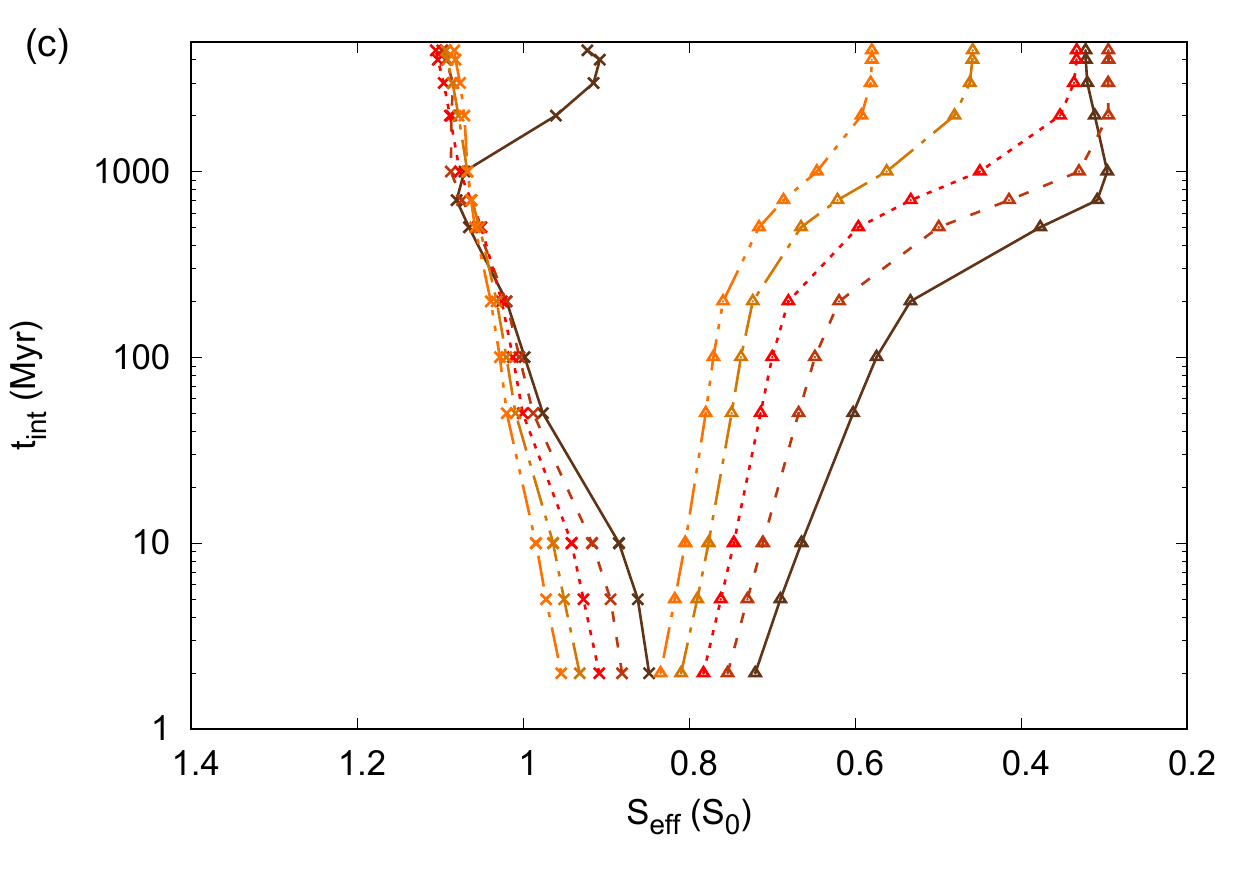}
    \includegraphics[width=0.4\textwidth]{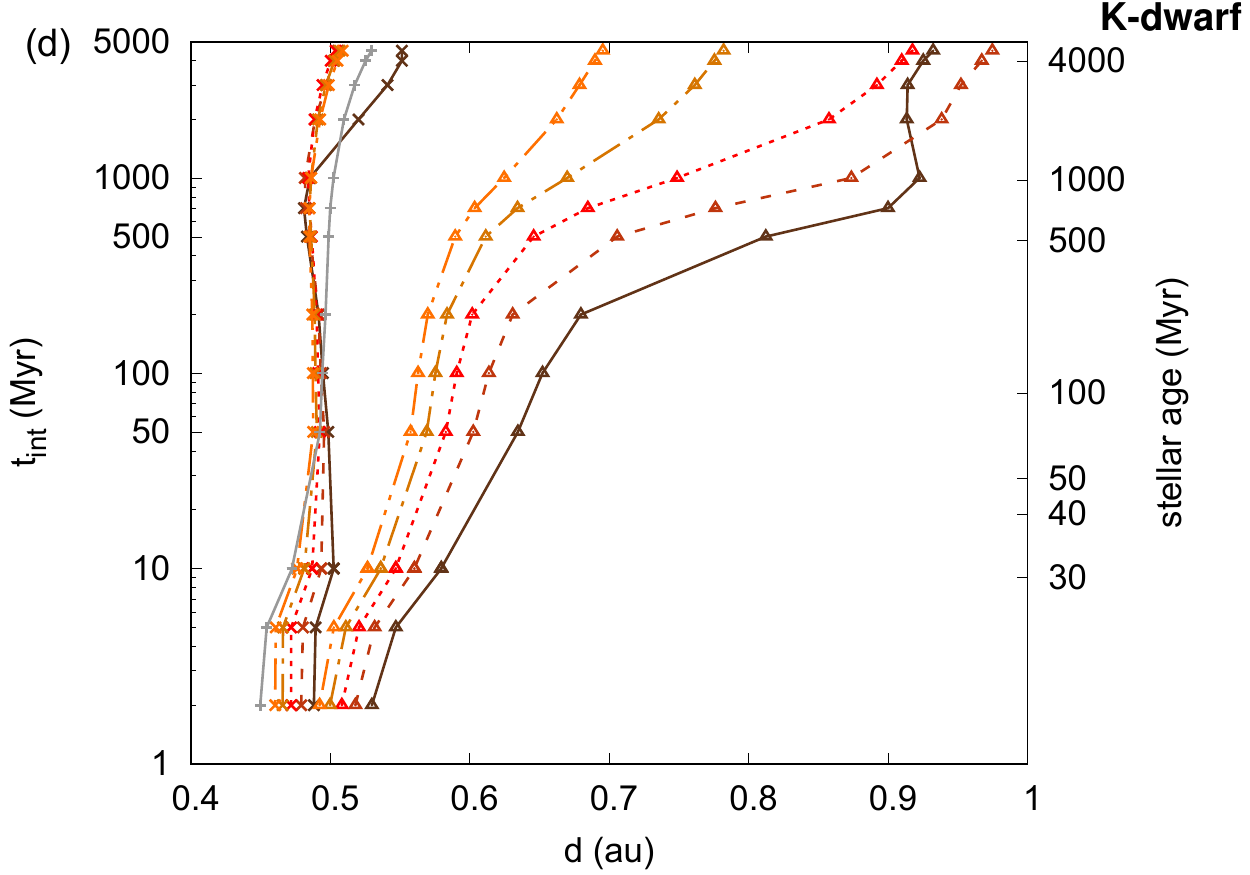}
    \includegraphics[width=0.4\textwidth]{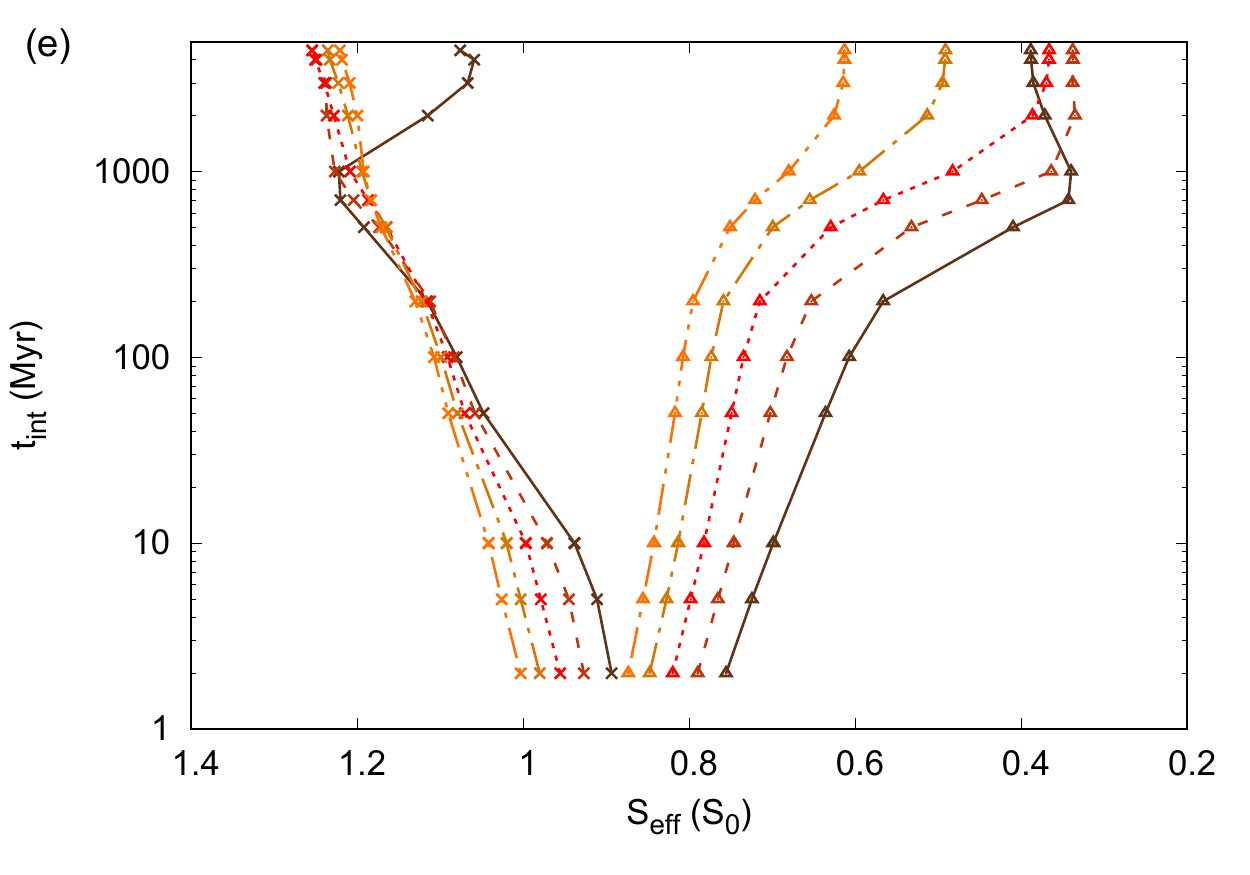}
    \includegraphics[width=0.4\textwidth]{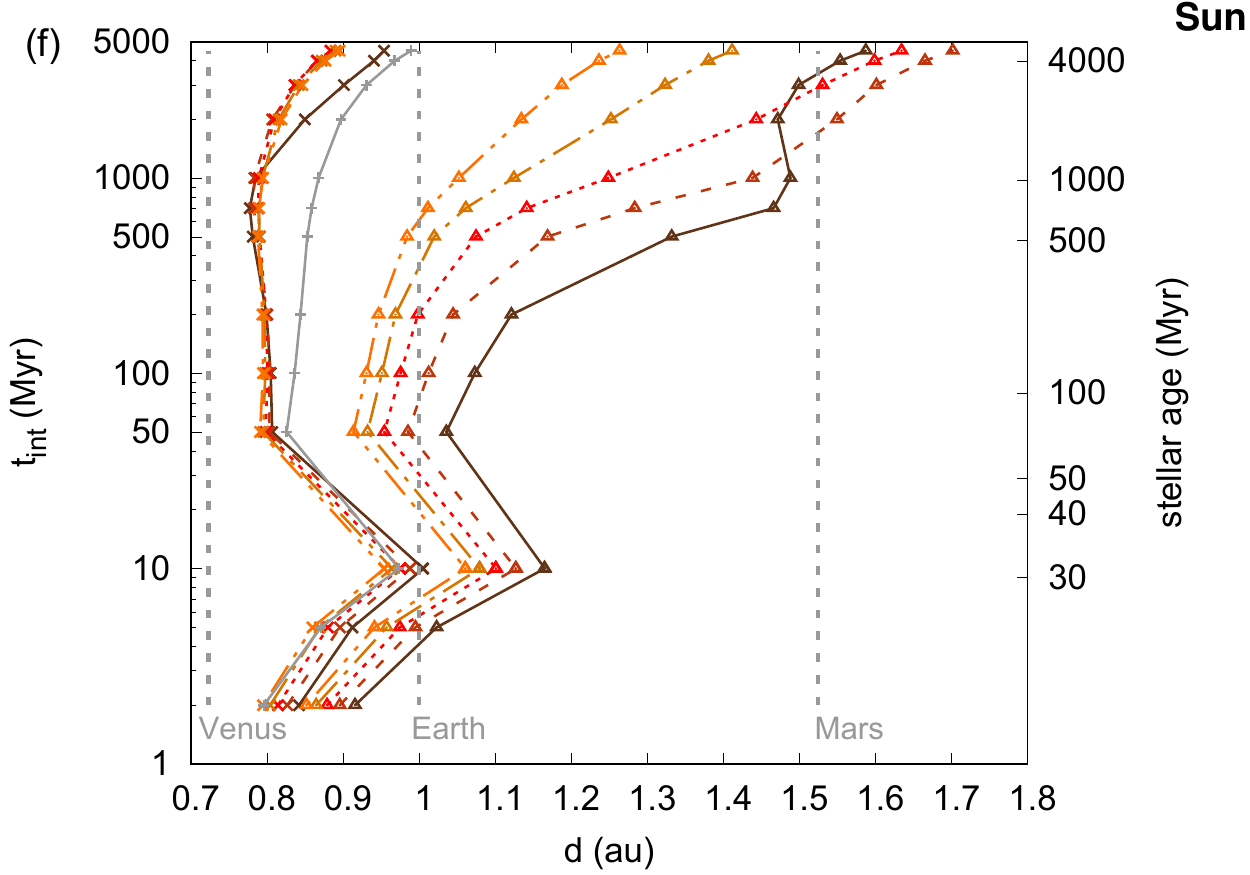}
    \includegraphics[width=0.4\textwidth]{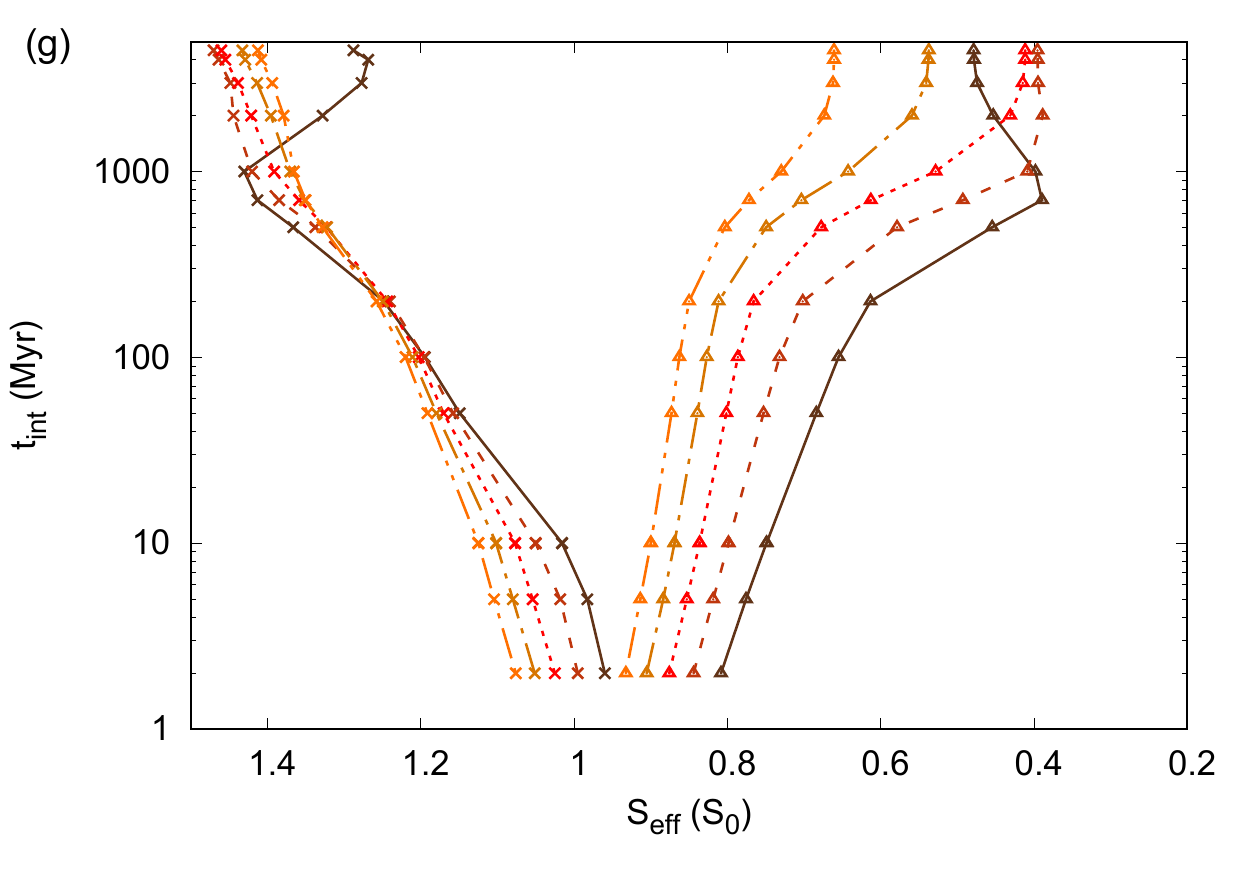}
    \includegraphics[width=0.4\textwidth]{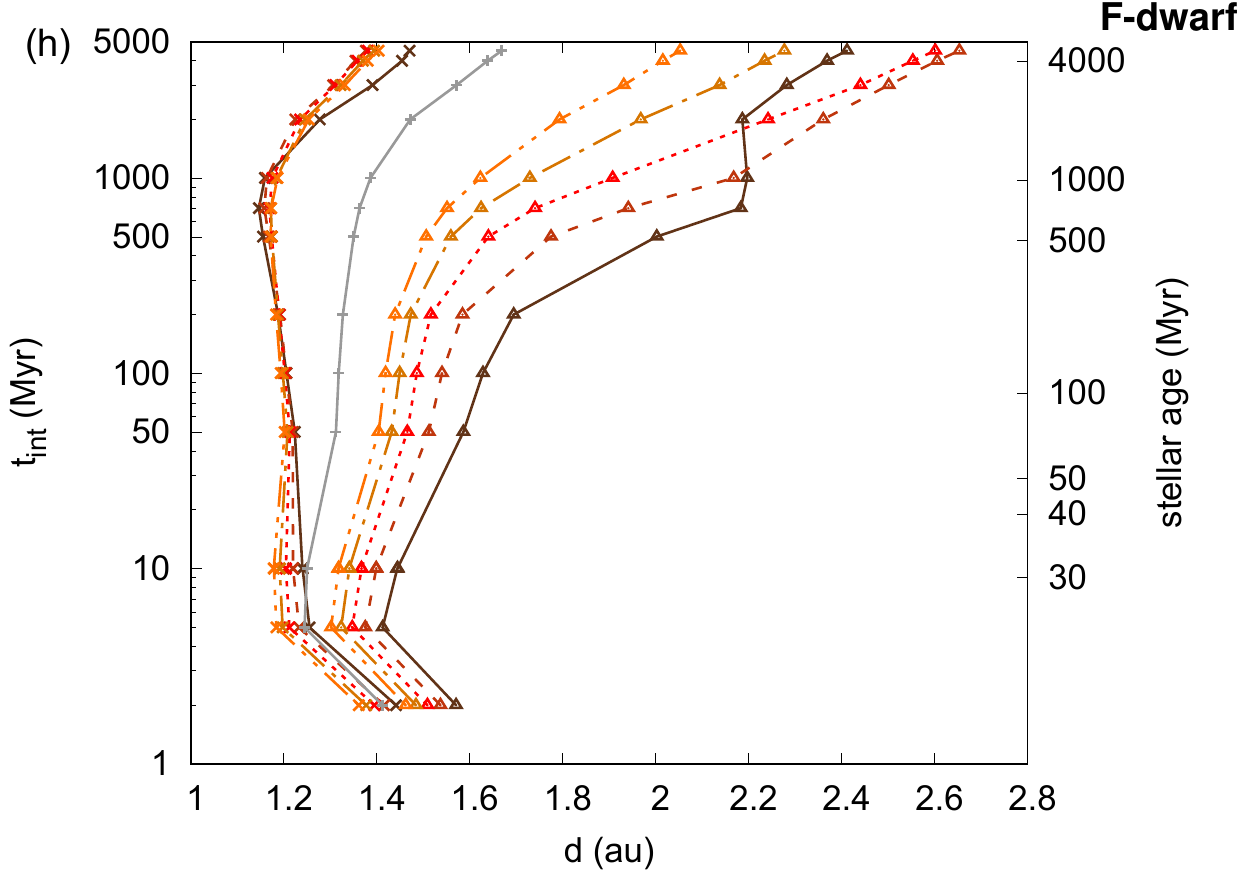}
    \end{center}
    \caption{HZ evolution for planets with an initial interior water content of 500\,ppm. The left column shows the habitable zone boundaries in terms of stellar irradiation (S$_0$), around an M, K, G and F-dwarf from top to bottom. The right column shows the HZ boundaries in orbital distance (au), which includes the stellar luminosity evolution. For illustration we indicate the orbital distance evolution of $S_\mathrm{eff}=1S_{0}$ with a grey solid line, which is the radiation the Earth receives from the Sun today, as well as the orbital distances of Venus, Earth and Mars in panel (f) with vertical grey dashed lines. The stellar age corresponding to the interior evolution times is indicated on the secondary y-axis where applicable.} 
    \label{fig:HZ_evolution_H500}
    \end{figure*}
    
Figure \ref{fig:HZ_evolution_H500} shows the evolution of the habitable zone boundaries of the stagnant-lid planets with an initial mantle water concentration of 500\,ppm for different oxygen fugacities and host stars. The left panel shows the evolution in terms of stellar irradiation showing the impact of the build-up of the atmospheres via outgassing, while the right panel shows the evolution in terms of orbital distance, which also includes the evolution of the stellar luminosity.

\subsubsection{HZ boundaries in terms of stellar irradiation }
In terms of stellar irradiation the habitable zone expands due to the build-up of H$_2$O and CO$_2$. 
For the planet around the M-dwarf AD Leo (Fig.~\ref{fig:HZ_evolution_H500}a) the shift of the inner habitable zone (IHZ) boundaries towards the star (to higher values of S$_{\mathrm{eff}}$) mainly occurs during the first $\approx$ 100\,Myr of secondary outgassing.
This timescale is shorter than that of the build-up of the entire atmospheric-oceanic water reservoir, which corresponds to about 500\,Myr (see dark blue line in Fig.~\ref{fig_h2o_CO2_out_IW0}a). The small amount of outgassed H$_2$O during this initial phase of $\approx$100\,Myr already causes the atmosphere to become opaque at similar pressure levels. Increasing the water reservoir increases the surface temperature but not the infrared radiation leaving the planet. This effect, called the radiation limit, has been widely discussed in the literature \citep[e.g.,][and references therein]{Nakajima1992,Abe2011,Goldblatt2012,Leconte2013} and is linked to the runaway greenhouse limit.  A closer look into \citet{Nakajima1992} (their Fig.~5a) indicates that the radiation limit is approached at temperatures around 350\,K at which the water vapor saturation pressure is about 0.4\,bar. The convergence to a constant outgoing long-wave radiation when moving from temperate to warmer climates at about 350\,K has also been found by \citet{Kasting1993hz} and \citet{Kopparapu2013}. 

For planets around the M-dwarf AD Leo the stellar irradiation needed to reach the IHZ boundary is lower than one solar constant ($S_0$), the amount of stellar irradiation the Earth receives from the Sun (Fig.~\ref{fig:HZ_evolution_H500}a). That is due to the fact that the M-star mainly radiates at longer wavelengths. This radiation is less efficiently scattered and more effectively absorbed by H$_2$O and CO$_2$ in the planetary atmosphere, leading to larger heating rates.
This has also been found by 3D climate models  for Earth-like rotation rates \citep[see e.g.][]{Shields2014,Wolf2017}. At smaller rotation rates, e.g.~due to tidal locking, 3D model calculations have found that a negative cloud feedback may allow for much higher irradiations at the inner edge of the habitable zone \citep[see e.g.][and Sec.~\ref{discussions}]{Yang2014, Kopparapu2016, Way2018}. 
For $f_\mathrm{O_2}$ at the IW or lower, results suggest that the IHZ boundaries lie at the same value of stellar irradiation, while for larger $f_\mathrm{O_2}$, especially at $\mathrm{IW}+1$, the IHZ moves outwards to lower values of stellar irradiation. This is due to the combined effect of increased atmospheric CO$_2$ and decreased atmospheric-oceanic water reservoir outgassed from the interior. For a higher partial pressure of CO$_2$, less H$_2$O can be outgassed -- about 5\,bar at $\mathrm{IW}+1$ compared to 10\,bar at $\mathrm{IW}-1$. This is because at higher pressures the saturation pressure of water in the melt is higher. As a consequence, H$_2$O tends to stay in the melt, as shown in \citet{Tosi2017}. Therefore, less stellar irradiation is needed to heat the planet with a dense atmosphere to temperatures at which the smaller water reservoir would reside within the atmosphere. 
The outer habitable zone (OHZ) boundaries move to lower irradiation with time for all $f_\mathrm{O_2}$, reaching their final value at about 2000--3000\,Myr. The higher the $f_\mathrm{O_2}$, the lower the irradiation needed to reach the OHZ surface temperature of 273.15\,K which is the temperature we assumed for the OHZ boundary. This is due to the larger greenhouse effect of the higher CO$_2$ partial pressures outgassed at larger $f_\mathrm{O_2}$. The atmospheric partial pressure of water is the same for all OHZ scenarios, as it is determined via the surface temperature, which is the same for all cases. For an $f_\mathrm{O_2}$ of $\mathrm{IW}+0.5$ and $\mathrm{IW}+1$, the values of the stellar irradiation at the OHZ are approximately the same, despite an increase in CO$_2$ from about 6 to 18 bars. Increasing the amount of CO$_2$ in the atmosphere increases the greenhouse effect, but also the scattering effect of the atmosphere. For M-dwarfs these two processes balance each other for these CO$_2$ partial pressures, while for the other stars, as discussed below, the scattering effect starts to dominate the temperature response at CO$_2$ partial pressures above about 7\,bar.

For the planet around the K-star, the IHZ boundary extends towards the star up to about 700\,Myr, and the irradiation to reach the IHZ is very similar for all oxygen fugacities except for $\mathrm{IW}+1$ (see Fig.~\ref{fig:HZ_evolution_H500}c). The IHZ limit is found at radiations larger than one solar constant, since the amount of radiation at wavelengths where Rayleigh scattering is important is larger than for the M-dwarf star, which shows IHZ boundaries at irradiations smaller than one solar constant. 
The OHZ boundaries extend outwards to lower values of stellar irradiation and reach their final value between 2000--3000\,Myr. For nearly all cases, the OHZ boundary is at lower values of stellar irradiation for higher $f_\mathrm{O_2}$, corresponding to larger amounts of CO$_2$ outgassed. For $\mathrm{IW}+1$, however, the OHZ boundary moves inwards at later evolutionary stages due to the increased scattering effect of the accumulated CO$_2$ molecules. The maximum greenhouse effect is obtained at about 1000\,Myr, at partial pressures of CO$_2$ around 7 bar.
The maximum greenhouse effect becomes more apparent for the planets around the K-dwarf and more massive stars, since the stellar spectral energy distributions shift towards shorter wavelengths. The maximum greenhouse effect is caused by an increase in scattering of stellar light by denser atmospheres, as shown by e.g.~\cite{Kasting1993hz} for CO$_2$-dominated atmospheres. A similar effect is also obtained for N$_2$-dominated atmospheres, as shown e.g. by \citet{Keles2018}. 

For the planet around the Sun, the HZ boundaries show a similar behaviour as for the K-type star (Fig.~\ref{fig:HZ_evolution_H500}e). However the IHZ boundaries are closer in, and move even closer to the star with time due to the increase in scattering, which is more pronounced for the Sun than for the K-type star. This also causes a stronger shift towards the star for the OHZ at high oxygen fugacity ($\mathrm{IW}+1$), although the maximum greenhouse effect is found at about 1000\,Myr as for the planets around the K-type star.

This trend is continued for the planets around the F-type star (see Fig.~\ref{fig:HZ_evolution_H500}g). For these scenarios the HZ has the largest extent in terms of stellar irradiation, while it shows the smallest extent for the planets around the M-type stars. This difference in extent is caused mainly by the increased scattering of stellar irradiation by the atmosphere for the hotter stars at the IHZ. Therefore, planets around F-dwarfs can be habitable for higher irradiations than planets around M-dwarfs . Increased scattering of stellar irradiation also occurs at the OHZ for F-dwarfs compared to M-dwarfs, however the difference in total irradiation needed to reach the OHZ is much smaller than that at the IHZ boundary, which dominates the width difference of the HZ. With respect to the different oxygen fugacities of the mantle, the extent of the HZ is largest for $\mathrm{IW}+0.5$ while it is smallest for $\mathrm{IW}-1$ and $\mathrm{IW}+1$.

\subsubsection{HZ boundary distances}

The right panels of Fig.~\ref{fig:HZ_evolution_H500} show the evolution of the habitable zone in terms of orbital distance, accounting for the luminosity evolution of the star.

For the planets around the M-dwarf AD Leo, the pre-main sequence phase with the higher luminosity leads to a large shift of the habitable zone boundaries from values of around 0.43 to 0.5\,au to smaller values of about 0.15 to 0.2\,au during the first 200\,Myr of evolution (Fig.~\ref{fig:HZ_evolution_H500}b). After that large shift, the influence of outgassing, which is much stronger than that of the later evolution of the star as indicated by the solid grey line, can be clearly identified. The largest extent of the HZ is again obtained for $f_\mathrm{O_2}=\mathrm{IW}+0.5$. The least wide extent is now found for the lowest value of $f_\mathrm{O_2}$ considered here ($\mathrm{IW}-1$).

The other stars also show a luminosity evolution as they make the transition from the pre-main sequence to the main sequence.
As stated in Sec.~\ref{scenarios} we have set a later start of the secondary outgassing evolution for the planets around these stars, i.e. at 21\,Myr, because accretion is thought to last longer around higher-mass stars \citep[e.g.][]{Raymond2007}. This leads to a shorter time span over which these early luminosity changes have an impact on the habitable zone boundaries. For the planets around the K-type stars only a small shift in the HZ of around 0.05\,au is caused by the stellar evolution within the first 10\,Myr (Fig.~\ref{fig:HZ_evolution_H500}d). For the planets around the Sun a back and forth shift of the HZ boundaries of up to 0.3\,au is found for the first 50\,Myr (Fig.~\ref{fig:HZ_evolution_H500}f). For the F-star the HZ moves inward by about 0.2\,au during the first 5\,Myr of evolution (Fig.~\ref{fig:HZ_evolution_H500}h).

At later stages, the trends in the evolution of the HZ boundaries with time and $f_\mathrm{O_2}$ are very similar to those in stellar irradiation (shown in the left column of Fig.~\ref{fig:HZ_evolution_H500}): OHZ boundaries move outwards for increased outgassing of CO$_2$, but they move back inwards at the highest $f_\mathrm{O_2}$ ($\mathrm{IW}+1$) considered due to increased scattering by the atmosphere for CO$_2$ partial pressures larger than those needed to obtain the  maximum greenhouse effect.
In addition, however, the inner and outer boundaries slowly move away from the star due to its brightening. This shift is largest for the planets around the F-type star and smallest for the planets around the K-type star, and seems to be absent for the planets around the M-dwarf. This is caused by the faster main sequence evolution of the more massive and hotter stars (e.g.~F and G-dwarfs), while the cooler low-mass stars (K and M-dwarfs) show only a small luminosity evolution on the main-sequence over the time span we consider here. This outward shift of the OHZ due to the brightening of the stars is strongest for $f_\mathrm{O_2}=\mathrm{IW}+0.5$. While for the M-star this shift is rather small in terms of orbital distance (from about 0.2 to 0.28\,au), it becomes much larger for the other stars with a shift from about 0.6 to 0.95\,au for the K-dwarf, from about 1 to 1.6\,au for the Sun, and more than 1\,au for the F-type star from about 1.5 to 2.6\,au.

Our modelling results for planets around the Sun (Fig.~\ref{fig:HZ_evolution_H500}f) suggest that an Earth-like planet with a stagnant lid could not be habitable at the orbital distance of Venus at any time, while a planet located at 1\,au as for the Earth could be habitable throughout nearly the entire evolution for all oxygen fugacities assumed here. An Earth-like stagnant-lid planet at Mars' orbit may be habitable at later evolutionary times (larger than about 1700\,Myr) for oxygen fugacities at the IW and higher.

\subsubsection{Continuous habitable zone}

For Earth there is evidence for liquid water which dates back to about 3800\,Myr ago \citep[see e.g.][]{Kasting1993}. Our results demonstrate that the habitable zone boundaries shift with time not only due to the evolution of the stellar irradiation but also because the atmosphere evolves with time. Hence whether or not a planet is habitable depends not only on the distance to its host star but also on its age as well as on its interior composition and its outgassing evolution, among other factors such as stellar metallicity, planetary rotation rate or water delivery. Imposing the requirement that the planet has liquid water for 3800\,Myr reduces the range of possible orbital distances. This range of orbital distances which allows for habitable surface conditions on an Earth-like planet over a long time span is known as the the continuous habitable zone (CHZ), a concept discussed e.g.~by \citet{Hart1979} and \citet{Kasting1993hz}.

\begin{figure}[ht!]
\includegraphics[width=0.5\textwidth]{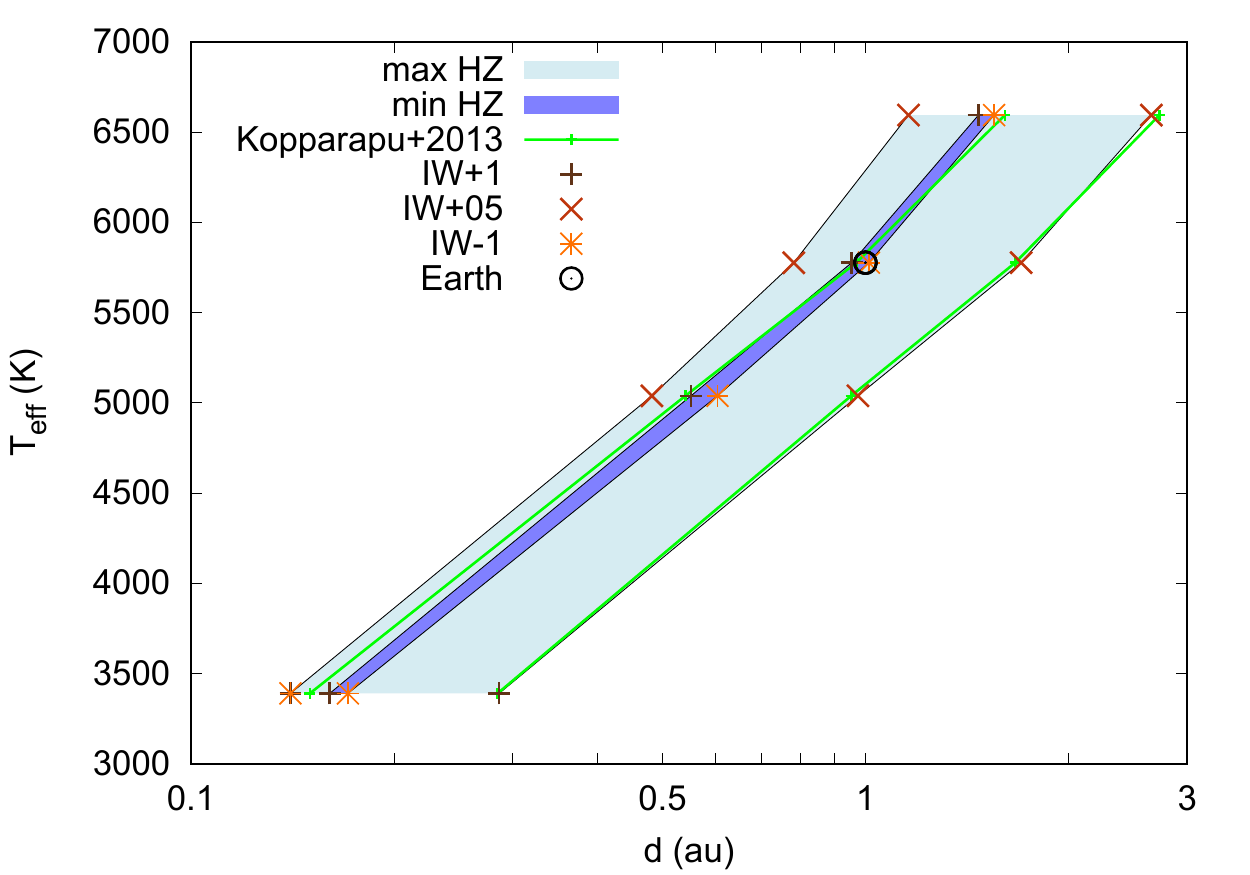}
\caption{Continuous HZ over 3800\,Myr with t$_{int}$ from 700 to 4500\,Myr and for an initial mantle water content of 500\,ppm. The minimum and maximum extent of the habitable zone are indicated by shadings. The oxygen fugacities responsible for the minimum and maximum extent of the CHZ are indicated by the colored symbols. The black circle shows the location of the Earth.}
\label{fig:HZ_continious}
\end{figure}

An aspect so far not considered when computing the CHZ is the impact of outgassing from the interior and the interior composition. Even when fixing the initial amount of water in the mantle, we obtain different widths of the HZ due to the different oxygen fugacities. Figure \ref{fig:HZ_continious} shows the minimum and maximum CHZ boundaries for the different host stars assuming an initial mantle water concentration of 500\,ppm. The maximum CHZ shows the range of orbital distances where an Earth-like stagnant-lid planet could be potentially habitable over 3800\,Myr from t$_{int}=$ 700 -- 4500\,Myr. The CHZ however depends on the oxygen fugacity of the mantle. For the boundaries of the minimum CHZ we set as a requirement that habitability is obtained for all of the oxygen fugacities assumed in this study (between $\mathrm{IW}+1$ and $\mathrm{IW}-1$) over 3800\,Myrs, which reduces the orbital distances covered drastically. We have compared the maximum CHZ boundaries with those of \citet{Kopparapu2013} and find a relatively good agreement, especially at the outer edge of the habitable zone. At the inner edges differences occur which are most probably due to the treatment of H$_2$O in the radiative transfer of the models, which is a major uncertainty in the calculation of the inner edge of the habitable zone as discussed by \citet{Yang2016}. Interestingly, the Earth resides just at the outer edge of the minimal CHZ.

\subsection{Habitability of stagnant-lid planets around M-dwarfs}
\label{Sec:M-Star_HZ}

For planets around M-dwarf stars the pre-main sequence phase with its much higher luminosities is thought to endanger planetary habitability even at later stages, as discussed e.g.~by \cite{Ramirez2014} and \cite{Tian2015}. Planets which reside within the HZ around their M-dwarf host star at later evolutionary stages, e.g.~at 4500\,Myr, have resided well inside the inner edge of the HZ during the pre-main sequence phase of their host star. Here we firstly show how the edges of the habitable zones evolve under the combined influence of stellar luminosity and outgassing evolution accounting for different M-dwarf classes and a variety of initial mantle water concentrations. Secondly, we discuss the possibility of rebuilding the water reservoir after severe water loss during the high-luminosity pre-main sequence phase. 
As already discussed in Sec.~\ref{Int_results}, for an initial water concentration of 34\,ppm we do not find any habitable surface conditions in the classical sense since the outgassed amount of water lies below its saturation pressure at 273.15\,K. We therefore analyze only cases where the outgassing is above 6.12\,mbar.

\begin{figure}[ht!]
\includegraphics[width=0.5\textwidth]{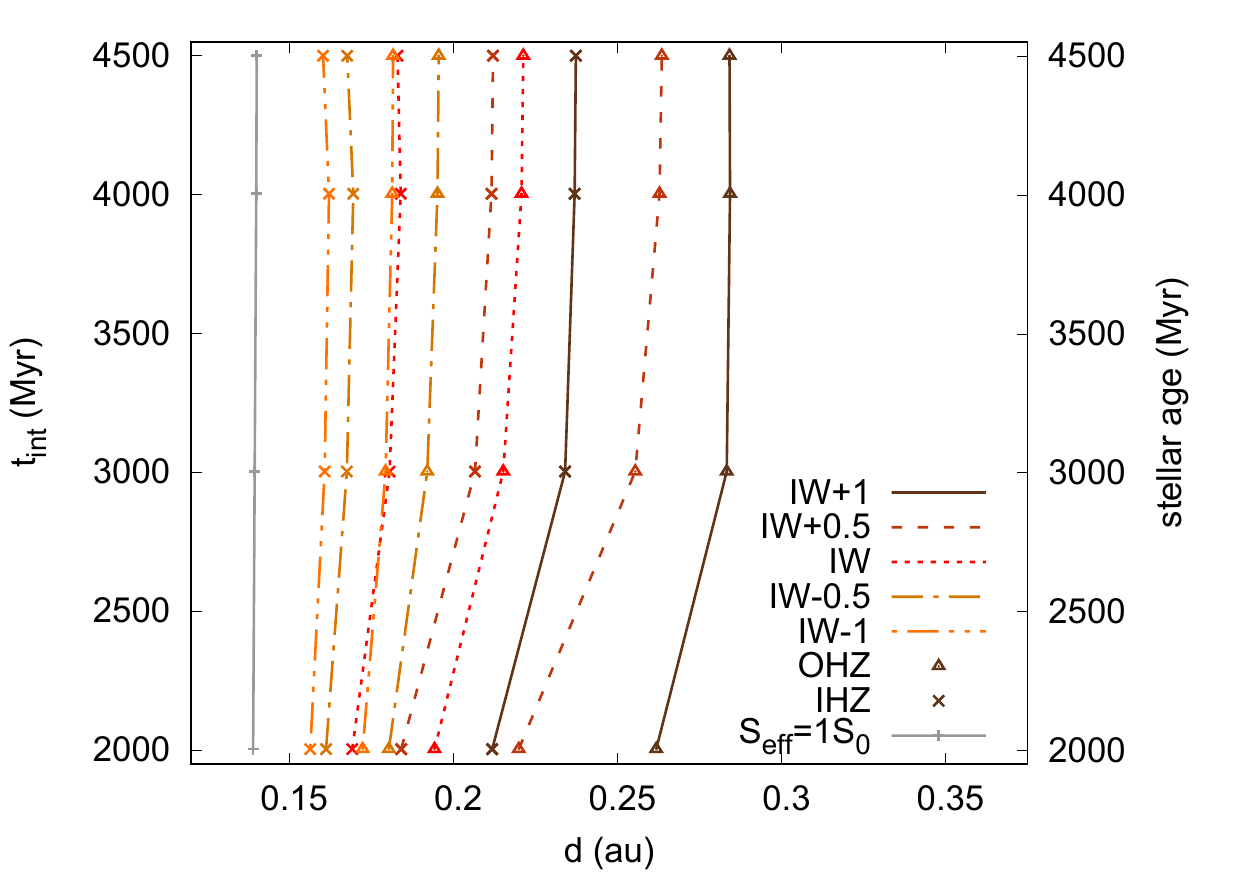}
\caption{HZ evolution of Earth-like stagnant-lid planets around the M-dwarf AD Leo for 62\,ppm initial water concentration in terms of orbital distance.}
\label{fig:ADL_HZ_H62_in_d}
\end{figure}

For 62\,ppm initial mantle water content there is a transition from outgassed water abundances below 6.12\,mbar for the first 1000\,Myr of evolution and above for later times (see Fig.~\ref{fig_h2o_CO2_out_IW0}a). The evolution of the habitable zone boundaries of a planet with this initial mantle water concentration around AD Leo is shown in Fig.~\ref{fig:ADL_HZ_H62_in_d}. It is shown starting from 2000\,Myr when the outgassed water reservoir is large enough to allow for liquid surface water. The boundaries of the habitable zone for the case of 62\,ppm initial water concentration in the mantle strongly depend on the amount of CO$_2$ outgassed. The boundaries of cases with little CO$_2$ in the atmosphere lie closer to the star, and there is no overlap in the habitable zones with low ($\mathrm{IW}-1$, $\mathrm{IW}-0.5$) and high ($\mathrm{IW}+0.5$, $\mathrm{IW}+1$) CO$_2$. While the stellar irradiation no longer changes at these later times, as can be inferred from the grey line showing $S_\mathrm{eff}=1S_0$, the HZ boundaries still evolve due to outgassing from the interior. This effect is most pronounced for the higher oxygen fugacities. 

\begin{figure*}[ht!]
\includegraphics[width=0.5\textwidth]{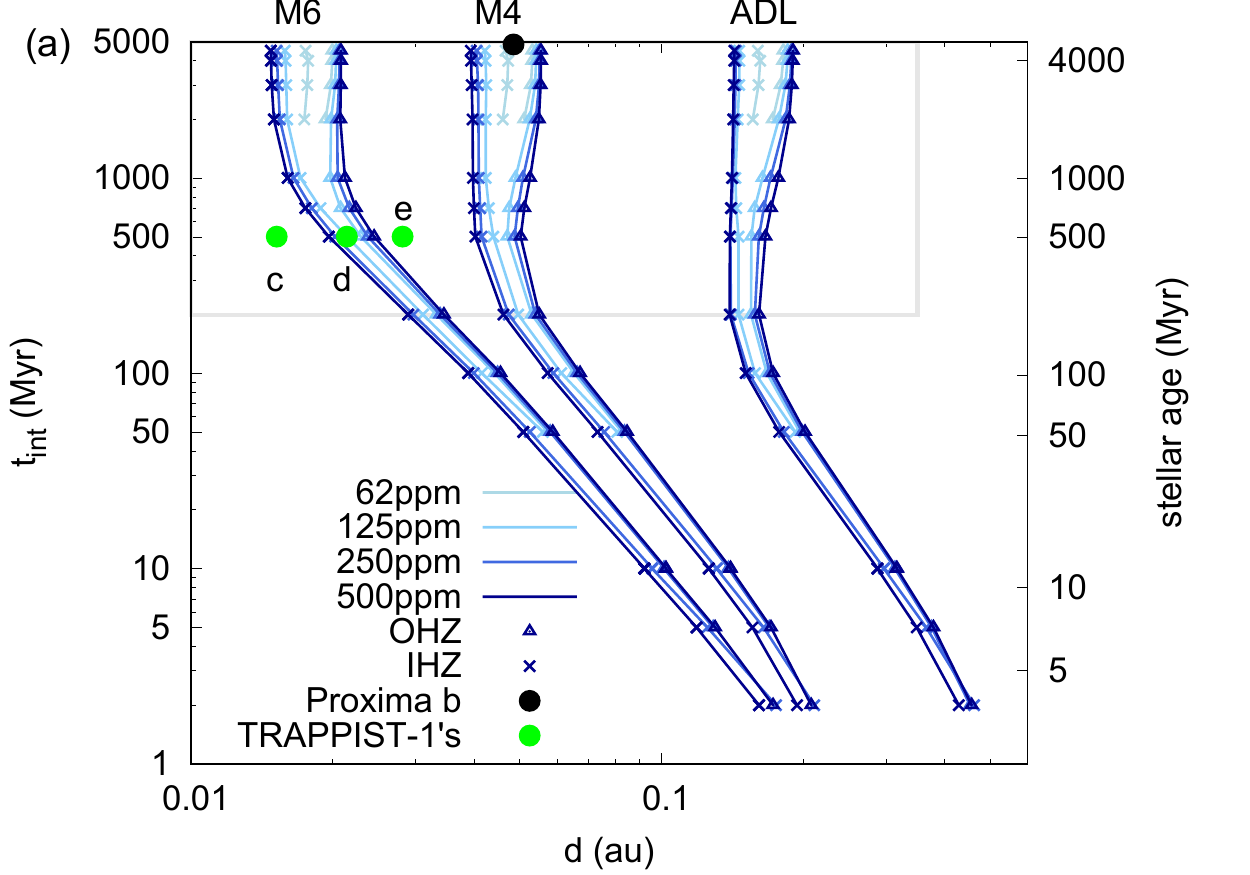}
\includegraphics[width=0.5\textwidth]{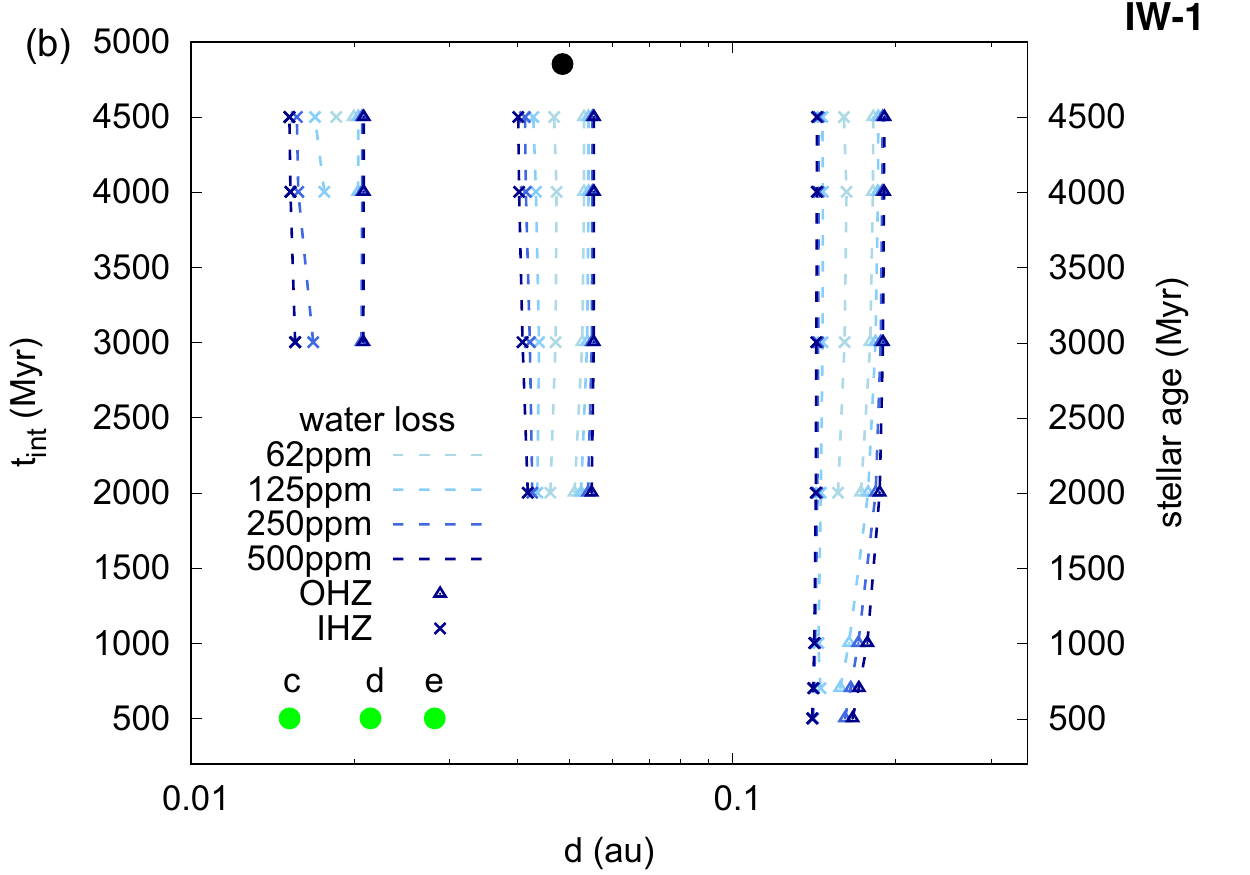}
\includegraphics[width=0.5\textwidth]{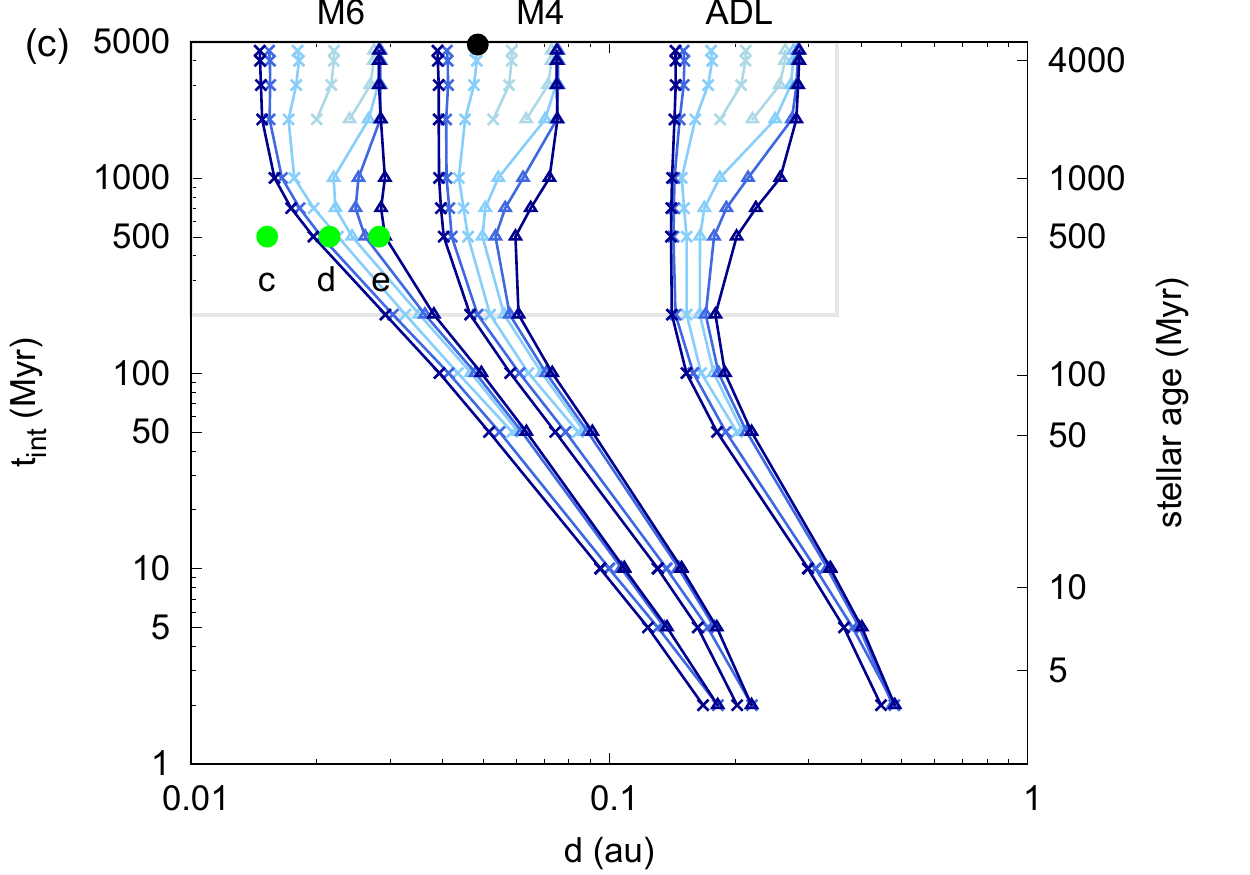}
\includegraphics[width=0.5\textwidth]{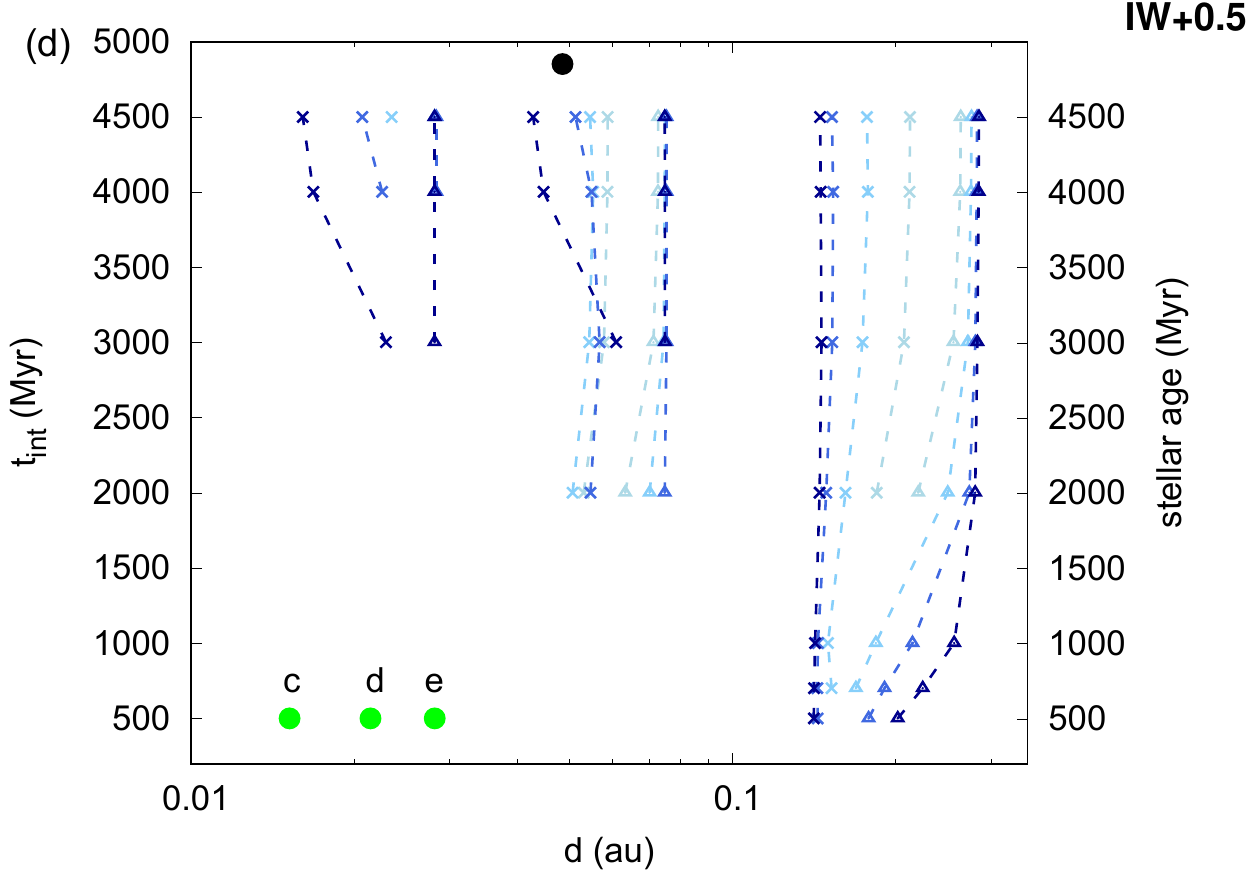}
\caption{HZ evolution for planets around the different M-dwarfs with different initial water contents (see legend) and oxygen fugacities, upper panels (a, b) $\mathrm{IW}-1$, lower panels (c, d) $\mathrm{IW}+0.5$ The left panel shows the HZ evolution for the full outgassed amounts of H$_2$O and CO$_2$. In the right column for each initial water concentration the HZ boundaries assuming that all water outgassed during the pre-main sequence phase of the star has been lost, are shown. For better visibility the y-axis has been changed to linear scale in the right panel. The locations of Proxima Centauri b and of TRAPPIST-1c, d and e are also shown. The grey box in the left panels indicates the time and distance range shown in the right panels.
}
\label{fig:MStars_HZ_d}
\end{figure*}

For initial mantle water concentrations of 125\,ppm and above the amount of water outgassed from the interior is almost always above 6.12\,mbar. Figure \ref{fig:MStars_HZ_d} shows the HZ boundary evolution in terms of orbital distance for the sample M-dwarf AD Leo, and the M4 and M6-dwarfs approximated by a Planck spectrum. The upper panels show the HZ evolution for an oxygen fugacity of $\mathrm{IW}-1$, the lower panels for an oxygen fugacity of $\mathrm{IW}+0.5$, leading to the HZs with the smallest and the largest extent in our study, respectively. From the left panels, which show the HZ boundary evolution over 4500\,Myr, it can be clearly seen that the time span during which the HZ boundaries shift due to the stellar evolution is shortest for the planets around AD Leo and longest for those around the M6-dwarf. While for the low value of $f_{\mathrm{O_2}}$ (Fig.~\ref{fig:MStars_HZ_d}a) the outward movement of the OHZ boundaries is rather small after the end of the pre-main sequence phase, for $\mathrm{IW}+0.5$ (Fig.~\ref{fig:MStars_HZ_d}c) a large outward movement is apparent. For $\mathrm{IW}+0.5$ the inner edge of the HZ also still shows some outward movement for the lower initial mantle water concentrations due to the build-up of CO$_2$ in the atmosphere. The spread of the HZ boundaries for the different initial mantle water abundances is larger for higher oxygen fugacities.

The long pre-main sequence phase of the M-dwarfs is believed to cause substantial atmospheric loss. This is justified not only because (young) M-dwarfs tend to be very active, but also because the bolometric luminosity is much higher. This suggests that a planet residing within the habitable zone around an M-dwarf at 4500\,Myr would be located well inside the inner habitable zone boundary at earlier evolutionary stages. The time span of this early atmospheric loss will strongly depend on the lifetime of the pre-main sequence phase. 
 
We find a strong impact of the stellar luminosity evolution on the HZ position up to about 200\,Myr for the planets around AD Leo (assumed to have a stellar mass of $0.4M_\mathrm{Sun}$), about 1000\,Myr for the planets around the M4-dwarf, and about 2000\,Myr for the planets around the M6-dwarf with a mass of $0.08M_\mathrm{Sun}$. The right panels (b, d) of Fig.~\ref{fig:MStars_HZ_d} show the HZ boundaries of the stagnant-lid planets assuming complete water loss during the pre-main sequence phase. Results are shown starting at 200\,Myr, which corresponds to a zoom into the region indicated by the grey box in the left panels (a, c). 
To estimate the maximum impact of severe water loss during the pre-main sequence phase, we here assume that all water outgassed during this phase (200\,Myr, 1000\,Myr, and 2000\,Myr for AD Leo, M4, and M6, respectively), has been completely lost. HZ boundaries are only shown if the atmospheric and oceanic water reservoir could be rebuilt by secondary outgassing after the pre-main sequence phase.
 
A HZ can be rebuilt for Earth-like stagnant-lid planets around AD Leo by secondary outgassing after the pre-main sequence phase. The HZ boundaries change marginally, mainly for $\mathrm{IW}+0.5$. The outgassed amount of water during the first 200\,Myr, hence the amount assumed to be lost completely, ranges from $1.25\times10^{-3}$ to 1.59\,bar. This makes up about 3 to 15\% of the total amount of water outgassed over the 4500\,Myr, which ranges from $4.02\times10^{-2}$ to 10.31\,bar. 

For Earth-like stagnant-lid planets around the M4 star with an oxygen fugacity of $\mathrm{IW}-1$ a HZ exists for all initial mantle water contents at times larger than 2000\,Myr, which is the first time step after imposing complete water loss at which we computed the atmospheres again. 
For an oxygen fugacity of $\mathrm{IW}+0.5$ the assumption of water loss shows a large impact on the position of the IHZ, especially for the planets with higher initial mantle water concentrations of 250 and 500\,ppm. For these two cases, 1.7 and 5.6\,bar are outgassed until 1000\,Myr, which makes up 83 and 78\% of the total amount outgassed over 4500\,Myr. The HZ boundaries calculated without water loss are no longer attainable via subsequent outgassing, hence water loss here leads to a decrease in HZ width.

For stagnant-lid planets around the M6 host star, the impact of the pre-main sequence on their habitability would be most severe, as this phase lasts about 2000\,Myr. For the cases with a low oxygen fugacity of $\mathrm{IW}-1$ and high initial mantle water concentrations of  250 and 500\,ppm the HZ can be restored. For an initial mantle water concentration of 125\,ppm it takes about 2000\,Myr to re-establish habitable conditions via outgassing, and for 62\,ppm only at 4500\,Myr habitability can be restored. For the higher oxygen fugacity of $\mathrm{IW}+0.5$ the water loss shows, as for the other two host stars, a much stronger impact. For an initial water concentration of 500\,ppm habitability is restored directly after the assumed water loss (here at the time step of 3000\,Myr), yet with a strong decrease in HZ width. However, the HZ widens again with time. For an initial mantle water concentration of 250\,ppm habitability is restored from 4000\,Myr on, while for 125\,ppm initial mantle water concentration habitability can be rebuilt only at 4500\,Myr. For 62\,ppm initial mantle water concentration no HZ can be rebuilt by 4500\,Myr. 

In Fig.~\ref{fig:MStars_HZ_d} we have also indicated the positions of Proxima Centauri b, and of three planets orbiting TRAPPIST-1. Proxima Centauri's values are similar to those we assumed for our M4-star, while TRAPPIST-1 has stellar parameters similar to those assumed for the M6-star. We can see that Proxima Centauri b \citep{Anglada-Escude2016}, with an orbital distance of 0.0485\,au and an estimated age of about 4800\,Myr \citep[from the age of $\alpha$ Centauri, see][]{Bazot2016}, lies in the range of the HZ boundaries for the planets around the M4-dwarf star. For most assumptions on the interior composition it lies within the HZ
even when assuming strong water loss. This is, however, not the case for an $f_{\mathrm{O_2}}$ of $\mathrm{IW}+0.5$ and initial water concentrations of 62\,ppm and 125\,ppm. For 250\,ppm the outgassing evolution after 4500\,Myr may allow Proxima b to reside within the HZ. For 500\,ppm initial water concentration secondary outgassing is large enough to rebuild an HZ after water loss during the pre-main sequence phase.

The orbital distances of the planets TRAPPIST-1 c, d, e \citep{Gillon2017} are indicated in green at 500\,Myr because it has been found that TRAPPIST-1 is at least 500\,Myr old. \citet{Burgasser2017} find an age of 7.6\textpm2.2\,Gyr for TRAPPIST-1. \citet{Bolmont2017} have studied the potential water loss of TRAPPIST-1b, c and d and found that these planets could have lost water reservoirs several times larger than one Earth ocean. Estimating the habitability at this stellar age would require further model calculations. At 500\,Myr an Earth-like planet at the position of TRAPPIST-1 d may be habitable for both oxygen fugacities depending on the initial mantle water concentration and water loss, see Fig.\ref{fig:MStars_HZ_d}. Planets at TRAPPIST-1 e's position could be habitable for $\mathrm{IW}+0.5$ and an initial mantle water concentration of 500\,ppm, if the water loss allows for a water reservoir larger than 6\,mbar.
At later stages, e.g., at 4000\,Myr, an Earth-like stagnant-lid planet at TRAPPIST-1 c's orbital distance may become habitable for low oxygen fugacities and large initial mantle water concentrations. While a planet at TRAPPIST-1 d's position would move outside the habitable zone for $\mathrm{IW}-1$, a planet at TRAPPIST-1 e's location may stay habitable over longer periods for $\mathrm{IW}+0.5$ and an initial mantle water concentration of at least 250\,ppm, even after severe water loss. The determination of the habitability of the detected planets in the TRAPPIST-1 system would however require additional calculations since they are probably substantially older than Earth and all have masses and radii different from Earth, which may lead to differences in the outgassing from the interior, as e.g. discussed by \citet{Noack2017} and \citet{Dorn2018}, and the atmospheric behaviour.

\section{Discussion}
\label{discussions}

For planets around M-Stars the long, highly luminous pre-main sequence phase has been suggested to endanger their habitability. However, some water reservoir may be rebuilt by secondary outgassing as shown in this paper. For low initial mantle water concentrations (34--125\,ppm) that may result from atmospheric escape during a long-term magma ocean phase \citep[see e.g.,][]{Hamano2013}, we found that the outgassing from the interior starts with a time-lag that facilitates rebuilding of an oceanic-atmospheric water reservoir. In the presence of a relatively dry mantle, in fact, the solidus temperature is high \citep[e.g.,][]{katz2003} and the mantle then needs 1000--2000\,Myr to heat up sufficiently for partial melting to occur and outgassing to be possible (see Fig.~\ref{fig:interior_evolution}). 

When assuming, as done here, that both the magma ocean and the accretion phases last about a million years for planets around M-dwarfs the pre-main sequence phase of early and mid M-dwarfs (here AD Leo and the M4-dwarf) ends at a time at which no outgassing from the interior occurs for the low initial water concentrations. Hence, the majority of H$_2$O and CO$_2$ is outgassed after the transition from the pre-main sequence to the main sequence phase, allowing for rebuilding a surface water reservoir.  
For the planet around the M6-star the pre-main sequence phase lasts longer, up to about 2000\,Myr. Hence outgassing from a water-poor interior already occurs at times when atmospheric loss should still be strong. Habitability may only be achieved for higher initial mantle water concentrations and at later evolutionary stages on these planets.\\

\noindent Influence of the magma ocean phase\\
Longer magma ocean phases would lead to a shift in the beginning of the secondary outgassing. For planets with low initial mantle water concentrations which show a delay in the outgassing, it may therefore be the case that the main phase of secondary outgassing begins after the end of the pre-main sequence phase. If the magma ocean phase is too long, however, atmospheric loss could deplete the interior mantle water reservoir too strongly.
We have evaluated the length of the magma ocean phases for the luminosity evolutions, by using the equations given in the supplementary material of \citet{Hamano2013}. They introduce two types of magma ocean phases, a long-term and a short-term magma ocean phase, the attainment of which depends on the host star's irradiation. If the irradiation is higher than the achievable outgoing longwave radiation of a water-dominated atmosphere, also known as the radiation limit \citep[e.g.,][]{Nakajima1992}, the planet is trapped in a long-term magma ocean which can only be exited via atmospheric water loss, or, in our case, by a decrease in stellar irradiation. Using Eq.~S12 in the supplementary material to \citet{Hamano2013}, we find at which evolutionary time the planets could switch from such a long-term to a short-term magma ocean.
For planets around AD Leo, which reside at the inner HZ boundary at 4500\,Myr, a switch from the long-term magma regime into a cooling regime would occur around 70\,Myr after the formation of the star. This transition would occur after about 600\,Myr for the planet around the M4-dwarf and after about 1800\,Myr for the planet around the M6-dwarf. Hence, especially planets around later M-dwarf stars may stay in a magma ocean phase over a very long time.
Secondary outgassing may therefore set in later than assumed in our study. The influence of such a longer magma ocean phase is hard to predict without detailed modelling. On the one hand water loss during a prolonged magma ocean may lead to a mantle too dry to obtain a large enough  oceanic-atmospheric water reservoir from secondary outgassing. On the other hand, if the stellar luminosity decreases quickly, the mantle may still be sufficiently wet to allow for a build-up of a water reservoir after the pre-main sequence phase via secondary outgassing.\\

\noindent Impact of stratospheric temperatures on the HZ\\
To estimate the impact of our assumptions about the stratospheric temperatures upon the width of the HZ, we have run additional calculations. We have compared the results for the outer HZ at 4500\,Myr with the different stratospheric temperatures, 150 and 200\,K,  for an oxygen fugacity of $\mathrm{IW}+0.5$ and an initial mantle water concentration of 500\,ppm and found differences in orbital distance of 0.04\,au for the M-dwarf AD Leo ($\approx$ 20\%) and 0.4\,au for the F-dwarf $\sigma$ Bootis ($\approx$18\%). At the inner edge of the habitable zone for the cases, the stratospheric temperatures assumed influence the results only for low water abundances, hence early evolutionary stages, since for a higher partial pressure of water the temperature profile is dominated by the convective lapse rate (see Fig.~\ref{fig:t_profiles}). Decreasing the stratospheric temperature to 150\,K leads to orbital distance differences of around 0.01\,au. Increasing the stratospheric temperature to 250\,K leads to larger orbital distance differences of up to 0.1\,au for a stagnant-lid planet with an oxygen fugacity of $\mathrm{IW}+0.5$ and an initial mantle water concentration of 500\,ppm around the F-dwarf $\sigma$ Bootis after 2\,Myr of interior evolution.
Comparing the effect of the stratospheric temperatures on the orbital distance of the HZ boundaries, to the effect which could arise from processes not included in this study, like e.g. clouds, other radiative species, the impact is minor.
\\

\noindent Extensions of the habitable zone\\
The amount of water outgassed from the interior can be very low for the planets considered here. Even for those with water partial pressures above 6.12\,mbar, i.e., those for which liquid water may be stable on the planetary surface for temperatures above 273.15\,K, the question remains as to what conditions would be present on such planets.

In our 1D calculations we assume that the atmospheres of our planets are saturated with water vapor \citep[as done in][]{Kasting1993hz, Kopparapu2013}, which leads to an IHZ that is located further away from the stars compared to 3D model results, which account for a hydrological cycle self-consistently. The drying of the atmosphere of a planet with a substantial water reservoir leads to the depression of the global mean relative humidity and hence allows for higher irradiations as shown by \citet{Abe2011,Leconte2013,Wolf2014,Popp2015}. The mean relative humidity has a large impact on the HZ boundaries \citep{Zsom2013} and surface temperatures \citep{Godolt2016}. The relative humidity of the atmospheres may additionally depend on the spectral energy distribution of the central star. Increased water concentrations have been obtained for planets around M and K-dwarf stars, partly caused by the higher portion of stellar NIR radiation \citep{Godolt2015,Fujii2017}.

If the water reservoir is small and water is transported towards the poles, it may be stored there. This can reduce the water vapor in the atmosphere, especially in the heated equatorial region, which then allows for a more efficient cooling of the planet, since the thermal radiation of the hot planetary surface is not blocked by the water vapor anymore and can leave to space. The outgoing long-wave radiation in this case is less limited than for planets with a large, planet-wide water reservoir \citep[see e.g.][]{Abe2011}. \citet{Kodama2018} have investigated for which water reservoir a transition to such a land planet regime may occur. Assuming zero obliquity and eccentricity, they find that Earth-sized planets with Earth-like topography and rotation rate become land planets if the water reservoir falls below 1\% of the mass of the Earth's oceans, i.e., about 2.7\,bar. The water is then located at higher latitudes, which are not in contact with the Hadley cell. 
For our scenarios, the amount of outgassed water implies that planets with an initial mantle water concentration below 250\,ppm would fall into the land planet regime for an Earth-like topography.

However, the rotation rate of the planet influences the meridional water transport. For rotational periods larger than that of the Earth, the Hadley cell extends over a wider range of latitudes \citep{DelGenio1987}. Planets around M-dwarfs may be tidally locked. Depending on whether other planets exist in the system which may excite a non-zero eccentricity \citep[see e.g.][]{Correia2008}, it may be locked into a synchronous orbit with a permanent day side or into another resonance. 
For tidally locked planets in synchronous orbit at the location of the inner HZ boundary at 4500\,Myr, the rotational periods would be 30.47 days for the theoretical planets around AD Leo, 7.8 days for the planets around the M4-dwarf, and 2.32 days for planets around the M6-dwarf. The study of \citet{DelGenio1987}, which assumed Earth-like planets with different rotational periods (without assuming a synchronous orbit) suggests Hadley cells which spread to latitudes larger than 80\textdegree\ for a rotational period of 30 days. For a rotational period of 7.8 days the Hadley cell would span to a latitude of about 60\textdegree\ and to about 30--40\textdegree\ of latitude for an orbital period of two days. Hence, this effect of storing water in the polar regions may be more effective for planets around the later M-dwarf stars, while for tidally locked planets around early M-dwarfs a giant Hadley cell may extend over the entire hemisphere. Hence, especially for the planets around the later M-dwarfs the storage of water at the poles may widen the HZ beyond the values we have computed here. 

For cases where the water reservoir is large enough to allow water at the substellar point, the inner edge of the HZ around M-dwarfs may be extended towards the star due to the build-up of an effectively scattering cloud deck \citep{Yang2013,Yang2014}. The strength of this effect however depends on the rotation rate \citep{Kopparapu2016}, as well as on the oceanic heat transport \citep[e.g.][]{Way2018} and will certainly also depend on the water reservoir. 
Hence, extensions of the inner HZ boundaries computed here are possible and should be investigated further, accounting for different sizes of water reservoirs and rotation rates.
Extensions beyond the outer edge of the HZ discussed here, have been proposed by studies including the effect of additional gases, such as molecular hydrogen (H$_2$) \citep[e.g.][]{Pierrehumbert2011H2} or methane \citep[e.g.][]{Ramirez2018b}. Especially the amount of primordial H$_2$ retained from the accretion phase plays an decisive role for planetary habitability in general, since these large hydrogen envelopes will likely render the planets uninhabitable, see e.g.~\cite{Owen2016}.\\

\noindent Potential sinks of CO$_2$\\
While we have estimated the effect of water for the inner HZ boundary by assuming that all water outgassed until the end of the pre-main sequence has been lost, we neglected the loss of atmospheric CO$_2$ to space in our study. Water loss mainly occurs over photodissociation of H$_2$O and subsequent loss of hydrogen to space. Losing molecules heavier than hydrogen is much more difficult due to the higher molar masses, but it may still be possible for strong irradiation.
The outgassing of CO$_2$ from the interior extends over longer time scales than that of water outgassing (see Sec.~\ref{Int_results}). For low initial mantle water concentrations the CO$_2$ outgassing also occurs after a time span of no outgassing due to the shift of the solidus to higher temperatures for low water concentrations in the mantle. Hence, also CO$_2$ outgassing can occur after potential substantial atmospheric loss during the pre-main sequence phase.

Another important sink for CO$_2$ is the loss of CO$_2$ by weathering of the surface. We neglected this effect in the present study as in \citet{Tosi2017}. 
\citet{Foley2018} study the potential impact of weathering for stagnant-lid planets. They find that for large enough planetary carbon reservoirs and radiogenic heating, weathering and outgassing can balance each other for 1000--5000\,Myr hence enabling habitable surface conditions. On the one hand, for large carbon reservoirs they show that weathering is supply-limited, i.e., the supply of fresh surface material is likely too small such that CO$_2$ would accumulate in the atmosphere, which leads to a Venus-like, uninhabitable hot climate. On the other hand, for low outgassing rates, limited by the carbon reservoir, the planet would likely exist in a snowball state. For the snowball state limit, they use outgassing values determined by \citet{Haqq-Misra2016} and \citet{Kadoya2014} as lower limits, which are equal to 10 and 100\% of present Earth's outgassing. 10\% of Earth's CO$_2$ outgassing flux as adopted by \citet{Foley2018} corresponds to $\sim 2.6 \cdot 10^{10}$ kg/yr. In our models, the rate of CO$_2$ outgassing depends both on the initial water concentration and on the assumed redox state. For 62 ppm water in the mantle and $f_{\mathrm{O}_2}$ at IW, the outgassing rate during the phase of melt production peaks at about $2.5\cdot 10^{9}$ kg/yr, while for $f_{O_2}$ at IW+1 it is one order of magnitude larger, and hence comparable with the lower limit mentioned above. \citet{Foley2018} discuss weathering for Earth-like water reservoirs. In our scenarios, the weathering of the surface could be limited by the  smaller surface water reservoir and lower precipitation rates. Therefore, although our outgassing rates are close to the lower limits adopted by \citet{Foley2018} that might lead to snowball states, the overall impact of weathering on our findings is not easily predictable and will require further study. 
\citet{Foley2018} furthermore consider metamorphic outgassing as an additional source of CO$_2$, which could increase CO$_2$ outgassing rates. Metamorphic outgassing refers to the release of CO$_2$ associated with the decarbonation of buried carbonated crust. On Earth, the contribution of this mechanism to the planet's long-term carbon cycle is thought to be potentially important but is still poorly understood \citep{evans2011}. Furthermore, the release of carbon due to metamorphic processes is generally associated with partial melting in subduction zones \citep[e.g.,][]{dasgupta2010} and deformation in active orogenic regions \citep[e.g.,][]{evans2008}, two settings that are intimately related to plate tectonics and hence absent in a stagnant-lid planet. Whether metamorphic outgassing on such bodies can be as relevant as on a planet with plate tectonics is therefore unclear and deserves more investigation.\\

\noindent Impact of the stellar irradiation\\
In this paper, we have considered the effect of an increased stellar luminosity during the pre-main sequence phase. So far, we neglected the change in the stellar spectral energy distribution and the impact of stellar activity. Especially late-type M-dwarfs may show increased stellar activity over a much longer time span than solar-like stars \citep[e.g.][]{Schneider2018}.  
This may increase the duration of atmospheric water loss. To determine the potential habitability for individual planets, such as Proxima Centauri b, or the planets around TRAPPIST-1, it is therefore crucial to account for all information available for the planets and the stars in order to estimate the impact of the individual processes and their interactions, \citep[e.g.][]{Meadows2018,Ribas2016,Turbet2016,Turbet2018}.

\section{Summary and Conclusion}
\label{summary}

We have studied the co-evolution of stagnant-lid Earth-like planets with their M, K, G, and F-dwarf host stars, accounting for the change in atmospheric mass and composition due to the evolution of the planetary interior and secondary outgassing of H$_2$O and CO$_2$ together with the luminosity evolution of the host star. 

We showed that the width of the habitable zone for stagnant-lid planets strongly depends on the outgassing of CO$_2$ from the interior, which is controlled by the oxygen fugacity. The classical HZ width can be obtained by secondary outgassing depending on the interior composition. The HZ may be much narrower for low oxygen fugacities or strong losses of CO$_2$ e.g. via weathering or atmospheric loss processes.

The initial mantle concentration of water is crucial for the outgassing history from the interior. For low initial mantle water concentrations a period of suppressed outgassing (of H$_2$O and CO$_2$) can occur. This effect may allow for the build-up of a secondary water reservoir by outgassing for planets around M-dwarfs after the high-luminosity pre-main sequence phase, which is thought to endanger their habitability.

\begin{acknowledgements}
The authors thank the anonymous referee for their helpful comments. M.~Godolt gratefully acknowledges support by the DFG through the project GO 2610/1-1 and the priority program SPP 1992 "Exploring the Diversity of Extrasolar Planets" (GO 2610/2-1). N.~Tosi acknowledges support by the Helmholtz association (VH-NG-1017) and by the DFG through the priority programs 1922 ``Exploring the diversity of extrasolar planets'' (TO 704/3-1) and 1883 ``Building a habitable Earth'' (TO 704/2-1). T.~Ruedas was funded by DFG grant RU 1839/2.
\end{acknowledgements}
    
\bibliographystyle{aa}
\bibliography{references}
\end{document}